\documentclass{aastex631}

\DeclareMathAlphabet\mathbfcal{OMS}{cmsy}{b}{n}
\usepackage{float}
\usepackage{hyperref}
\usepackage{amsmath}
\usepackage{url}

\usepackage{multirow}
\usepackage{graphicx}
\usepackage{booktabs} 

\begin{document}

\title{DART-Vetter: A Deep LeARning Tool for automatic triage of exoplanet candidates}

\correspondingauthor{Stefano Fiscale}
\email{stefano.fiscale001@studenti.uniparthenope.it}

\author[0000-0001-8371-8525]{Stefano Fiscale}
\affiliation{UNESCO Chair ``Environment, Resources and Sustainable Development'', Department of Science and Technology, Parthenope University of Naples, Italy;}
\affiliation{Department of Science and Technology, Parthenope University of Naples,  Centro Direzionale di Napoli, Naples, I-80143, Italy.}
\affiliation{INAF, Osservatorio Astronomico di Capodimonte, Salita Moiariello, 16, Naples, I-80131, Italy.}

\author[0000-0002-0271-2664]{Laura Inno}
\affiliation{Department of Science and Technology, Parthenope University of Naples,  Centro Direzionale di Napoli, Naples, I-80143, Italy.}
\affiliation{INAF, Osservatorio Astronomico di Capodimonte, Salita Moiariello, 16, Naples, I-80131, Italy.}

\author{Alessandra Rotundi}
\affiliation{UNESCO Chair ``Environment, Resources and Sustainable Development'', Department of Science and Technology, Parthenope University of Naples, Italy;}
\affiliation{Department of Science and Technology, Parthenope University of Naples,  Centro Direzionale di Napoli, Naples, I-80143, Italy.}

\author{Angelo Ciaramella}
\affiliation{Department of Science and Technology, Parthenope University of Naples,  Centro Direzionale di Napoli, Naples, I-80143, Italy.}

\author{Alessio Ferone}
\affiliation{Department of Science and Technology, Parthenope University of Naples,  Centro Direzionale di Napoli, Naples, I-80143, Italy.}

\author[0000-0001-6343-4744]{Christian Magliano}
\affiliation{INAF, Osservatorio Astronomico di Capodimonte, Salita Moiariello, 16, Naples, I-80131, Italy.}
\affiliation{Department of Physics “Ettore Pancini”, University of Naples Federico II, Naples, Italy.}

\author[0000-0001-8266-0894]{Luca Cacciapuoti}
\affiliation{European Southern Observatory, Karl-Schwarzschild-Strasse 2 D-85748 Garching bei Munchen, Germany.}

\author{Veselin Kostov}
\affiliation{NASA Goddard Space Flight Center, 8800 Greenbelt Road, Greenbelt, MD 20771, USA}
\affiliation{Citizen Scientist, Planet Patrol Collaboration, Greenbelt, MD, 20771, USA}

\author{Elisa V. Quintana}
\affiliation{NASA Goddard Space Flight Center, 8800 Greenbelt Road, Greenbelt, MD 20771, USA}

\author[0000-0002-2553-096X]{Giovanni Covone}
\affiliation{INAF, Osservatorio Astronomico di Capodimonte, Salita Moiariello, 16, Naples, I-80131, Italy.}
\affiliation{Department of Physics “Ettore Pancini”, University of Naples Federico II, Naples, Italy.}
\affiliation{INFN section of Naples, Via Cinthia 6, 80126, Napoli, Italy}

\author[0000-0002-4430-5135]{Maria Teresa Muscari Tomajoli} \affiliation{UNESCO Chair ``Environment, Resources and Sustainable Development'', Department of Science and Technology, Parthenope University of Naples, Italy;}
\affiliation{Department of Science and Technology, Parthenope University of Naples,  Centro Direzionale di Napoli, Naples, I-80143, Italy.}

\author[0009-0000-9738-0641]{Vito Saggese}
\affiliation{Department of Physics “Ettore Pancini”, University of Naples Federico II, Naples, Italy.}

\author{Luca Tonietti}
\affiliation{UNESCO Chair ``Environment, Resources and Sustainable Development'', Department of Science and Technology, Parthenope University of Naples, Italy;}
\affiliation{Department of Science and Technology, Parthenope University of Naples,  Centro Direzionale di Napoli, Naples, I-80143, Italy.}
\affiliation{INAF, Osservatorio Astronomico di Capodimonte, Salita Moiariello, 16, Naples, I-80131, Italy.}
\affiliation{Department of Biology, Federico II University of Naples, Naples, Italy.}

\author{Antonio Vanzanella}
\affiliation{National centre for Nuclear Research, Pasteura 7, 02-093, Warsaw, Poland}

\author{Vincenzo Della Corte}
\affiliation{INAF, Osservatorio Astronomico di Capodimonte, Salita Moiariello, 16, Naples, I-80131, Italy.}

\begin{abstract}

In the identification of new planetary candidates in transit surveys, the employment of Deep Learning models proved to be essential to efficiently analyse a continuously growing volume of photometric observations. To further improve the robustness of these models, it is necessary to exploit the complementarity of data collected from different transit surveys such as NASA's \textit{Kepler}, Transiting Exoplanet Survey Satellite (TESS), and, in the near future, the ESA PLAnetary Transits and Oscillation of stars (PLATO) mission.
In this work, we present a Deep Learning model, named \texttt{DART-Vetter}, able to distinguish planetary candidates (PC) from false positives signals (NPC) detected by any potential transiting survey. \texttt{DART-Vetter} is a Convolutional Neural Network that processes only the light curves folded on the period of the relative signal, featuring a simpler and more compact architecture with respect to other triaging and/or vetting models available in the literature. 
We trained and tested \texttt{DART-Vetter} on several dataset of publicly available and homogeneously labelled TESS and \textit{Kepler} light curves in order to prove the effectiveness of our model.
Despite its simplicity, \texttt{DART-Vetter} achieves highly competitive triaging performance, with a recall rate of 91\% on an ensemble of TESS and \textit{Kepler} data, when compared to \texttt{Exominer} and \texttt{Astronet-Triage}.
Its compact, open source and easy to replicate architecture makes \texttt{DART-Vetter} a particularly useful tool for automatizing triaging procedures or assisting human vetters, showing a discrete generalization on TCEs with Multiple Event Statistic (MES) $>$ 20 and orbital period $<$ 50 days.

\end{abstract}

 
\section{Introduction} \label{sec:intro}

Among the various methods used for discovering exoplanets, the transit technique from space telescopes has emerged as the most productive, contributing to the discovery of over three-quarters of the $\sim$5,800 currently confirmed exoplanets\footnote{\url{https://exoplanetarchive.ipac.caltech.edu/docs/counts_detail.html}}. 
Considering that a little more than thirty years ago, no exoplanet had been detected yet, this represents tremendous progress, that has been especially boosted by the two dedicated NASA's space missions: \textit{Kepler} \citep{2010ApJ...713L..79K} and its successor, the Transiting Exoplanet Survey Satellite (TESS) \citep{2014SPIE.9143E..20R}.

Since its launch in 2009 and throughout its nine years in space, \textit{Kepler} detected more than 34,000 events \citep{2016AJ....152..158T} in the light curves of over 150,000 stars. Among these events, about 4,700 were dispositioned as candidate planets and nearly 2,700 of them turned out to be confirmed as planets \citep[see e.g.][]{2016ApJ...822...86M, 2016PASP..128g4502M, 2018ApJS..235...38T}. TESS, launched in 2018, is actively scanning almost the entire sky - divided into 26 sectors - every two years by using a wide-field strategy. Each sector is observed for about 27 days, allowing TESS to capture high-precision brightness measurements of approximately six million stars. As of today, TESS detected more than 7,500 candidate planets, 622 of which have been confirmed as true planets\footnote{These numbers were retrieved from the \href{https://exoplanetarchive.ipac.caltech.edu}{NASA Exoplanet Archive} on May 2, 2025.}.

This large gap between the amount of collected data and the number of exoplanets found is because of the intervening time-consuming process that allows to rule out common false positives scenarios and identify potential exoplanet candidates, called $vetting$. 
Vetting consists in carefully examining and validating the detected transits to distinguish  transiting planets from astrophysical (e.g., eclipsing binaries or stellar variability) or instrumental (e.g., spacecrafts momentum dumps) false positives. 

While the exact analysis steps needed to promote an observed transit to a candidate and, possibly, a confirmed planet might vary depending on the instrument and detection method adopted, the team performing it, and the specific system under analysis, the basic steps for vetting transit signals are common.
Initially, vetters focus on the light curve data in which a transit-like feature has been detected, by using all available information (e.g. transit shape, difference between primary and secondary eclipses, etc.) to rule out false positives. At this stage, it is possible to discard clear false positives for which further analysis will not need to be carried out. This operation is often referred to as $triage$.

For candidates surviving triage, mostly transiting planets and eclipsing binaries, additional observations are collected and used to perform an independent validation or clearing out specific false positive scenarios. 
To remove false positives scenario such as background eclipsing binaries, a pixel-level analysis - based on the examination of difference images or centroids offset for the target stars - is performed \citep{bryson2013identification,hadjigeorghiou2024positional}. When such ancillary data are not available or not enough, statistical tools, as employed by \citet{10.1093/mnras/stac3404}, can be used to validate transits \citep{2015ascl.soft03011M,2021AJ....161...24G}. These tools assess the significance of the observed transit signal and provide a quantitative estimation of validation confidence.
Finally, the confirmation of the exoplanet identification is achieved through the radial velocity detection of its mass.
Whilst the last two steps in the process are strongly human-dependent, the first two, i.e. triage and vetting, can be automatized to some extent and several authors proposed automated procedures that can support the validation process at different stages.

Initial works focused on analyzing \textit{Kepler}'s dataset in order to produce a binary classification of transit events in planet candidates and false positives. Noteworthy examples include the decision-tree-based machine-learning codes such as \texttt{Autovetter} \citep{2015ApJ...806....6M}, \texttt{Robovetter} \citep{Coughlin_2016} and \texttt{SIDRA} \citep{2016MNRAS.455..626M}.
However, as more and more diverse data were acquired with the advent of new missions, such as K2 \citep{2014PASP..126..398H} and TESS, the complexity of the classification problem grew and deeper model started to become  appealing . \citet[][hereinafter SV18]{2018AJ....155...94S} introduced \texttt{Astronet}, a Convolutional Neural Network (CNN) designed for vetting planet candidates identified by the \textit{Kepler} mission but potentially applicable to other surveys as well.  
Indeed, \texttt{Astronet} and similar architectures have been applied to new datasets produced by surveys as K2 \citep{2019AJ....157..169D}, WASP \citep{2019MNRAS.483.5534S}, and NGTS \citep{2018MNRAS.478.4225A,2019MNRAS.488.5232C}. Adjustments to the methodology, incorporating new input information and refining data representation \citep{2018ApJ...869L...7A,jara2020multiresolution,2022ApJ...926..120V, 2023AJ....166...28V, 2024AJ....167..180L}, have resulted in enhanced classification performance. 

The CNN by SV18 was later customized for TESS by \citet[][hereinafter YU19]{2019AJ....158...25Y}.  
They obtained good performance for one of their trained models specialized on discarding all non-astrophysical signals, called \texttt{Astronet-Triage}, which was then incorporated into the TESS Quick-Look Pipeline (QLP) since 2019, as outlined by \citet{2021ApJS..254...39G}. 

Recently, \citet[][hereinafter TEY23]{2023AJ....165...95T} found that the screening process performed with YU19 model led to the exclusion of a significant number of potentially valid planet candidates. 
Hence, they proposed a new model, called \texttt{Astronet-Triage-v2}, which is trained to identify five different classes, including the one of interest, ``eclipsing signals'', in order to preserve more information for the next steps. 

To achieve this goal, they had to further increase the complexity of the model, providing seven different representations of the light curve and additional scalar information as input to the network. 
Whilst this approach produces higher performance and hence effectively reducing the number of  ``lost'' planets, it has some drawbacks: a) an increased complexity of the model that makes it more difficult to apply and test on other datasets, and b) a final classification that requires further interpretation or considerations.
Concerning the issue of model complexity \citep{hu2021model}, recent studies by \citet{2022JAI....1150011V,2022A&C....4100654V} have shown that increasing the number of trainable parameters of neural networks can actually reduce its predictive performance. 
The authors of these studies presented \texttt{Genesis}, a CNN similar to those presented in SV18 and \citet{2018ApJ...869L...7A} but with much less trainable parameters (or weights), and demonstrated that their simpler model performs as good as its more complex counterparts. 
This is an interesting notion, as reducing the complexity of a Deep Learning model can improve the generalization capability, prevent overfitting and being more easily exportable in order to be verified and used by different teams \citep{fiscale2025detection}. \\

In this work, we present a simplified CNN - designed to be easily replicable - to perform the initial sorting of signals into planet (PC) and not-planet candidates (NPC), named \texttt{DART-Vetter}.
In fact, based on the results from \texttt{Genesis}, we explored the opportunity to further simplify the architecture with respect to state-of-the-art CNNs used to perform triaging and vetting, and found that we can
significantly reduce both the input dimensionality \citep{Ferone2017116} and the number of layers without degrading the model performance. Overall, \texttt{DART-Vetter} shows robust predictive performance on TESS planetary candidates, while good generalization capabilities are achieved on \textit{Kepler} datasets for TCEs with Multiple Event Statistic (MES) $>$ 20 and orbital periods $<$ 50 days. The MES estimates the signal-to-noise ratio (S/N) of the signal against the measurement noise.

Our paper is organized as follows: In Section~\ref{sec:data}, we explain how light curve data from the \textit{Kepler} and TESS mission are collected, automatically processed and searched for potential events, that we use to prepare the input. 
In Section~\ref{sec:tce_and_labels}, we detail the transit signals and corresponding light curves used as the training and test set for our model \texttt{DART-Vetter}, along with the pre-processing steps used to standardize them in Section~\ref{sec:data_prep}. The architecture of the neural network and the training process are covered in Section~\ref{sec:model_architecture}. 
We present the quantified results of our classifier in Sections~\ref{sec:test} and \ref{sec:test_comparison} and discuss the implications in Section~\ref{sec:limitations}. Conclusions are drawn in Section~\ref{sec:conclusion}.

\subsection{Contributions/Novelty of This Work}

The approach we present with the Deep Learning-based model \texttt{DART-Vetter} offers several advantages in the field of exoplanets detection, as summarized below:

\begin{itemize}
    
    \item Faster training and TCEs predictions: we extensively employ dropout \citep{srivastava}, a regularization technique that not only mitigates overfitting and underfitting issues but also enables training and assessing an exponential number of network architectures in a computationally inexpensive manner. In contrast to other model selection methods such as $K$-fold cross-validation, dropout allows the model to be trained only once, providing faster performance and other advantages discussed in Section \ref{sec:model_training}.
    
    \item Balancing adaptability and reliability: the use of a compact input representation facilitates both data collection and preparation pipelines, and the application of our model on cross-mission data. We demonstrate in Section \ref{sec:test_comparison_comparison} that \texttt{DART-Vetter} can achieve competitive triage capabilities on \textit{Kepler} and TESS data. However, this versatility comes with limitations: since the model relies on a reduced set of input features, it shows reduced performance when classifying \textit{Kepler} TCEs with MES $<$ 20 or with orbital periods longer than 50 days as discussed in Section \ref{sec:test_comparison_snr}.
    
    \item Better generalization with reduced model complexity: with only 527,329 parameters, compared to state-of-the-art models with more than 100 million, \texttt{DART-Vetter} is significantly more compact. This reduction in complexity enhances generalization on unseen TCEs, minimizes the risk of overfitting on small datasets, and maintains competitive F1-scores across multiple datasets. We refer the reader to Section \ref{sec:model_architecture} and Section \ref{sec:test_comparison_rationale} for a more detailed discussion about the importance of reducing model complexity.
    
    \item Multi-class classification capability: although initially designed for binary classification, \texttt{DART-Vetter} can be quickly adapted to perform multi-class classification. As detailed in Section \ref{sec:test_multi}, this expands its applicability beyond binary classification tasks.
    
\end{itemize}

These features make \texttt{DART-Vetter} a versatile and easily reproducible tool that can ensure a correct minimization of the fraction of misclassified planetary transits.

\section{From light curves to TCEs}
\label{sec:data}
The data used in this work are light curves in which a \textit{threshold-crossing event} (TCE), i.e. a potential transiting planet signature with a minimum S/N of 7, is detected \citep{christiansenkepler,2002ApJ...564..495J}. 

In this work, we aim at building a model potentially able to automatically classify all TCEs detected in the light curves collected by a generic transiting survey. 
As we want to develop a survey-independent approach, we adopted all labelled TCEs publicly available from both \textit{Kepler} and TESS missions in order to obtain the largest and most diverse possible training set. 
As data can be collected and processed differently depending on the survey characteristics and data flow, we summarize here for the reader convenience how TCEs are detected in both \textit{Kepler} and TESS data. 

\subsection{Kepler Data}
\label{sec:kepler_data}
The \textit{Kepler} spacecraft collected photometric observations for $\sim$156,000 target stars at 29.4 min cadence (long cadence) and for 512 target stars at a shorter cadence of 58.85 $s$ \citep{2010ApJ...713L..79K}. 
These observations were divided into quarters, each lasting approximately 93 days. In each quarter, the telescope focused on a specific field of view in the sky, continuously monitoring the brightness of all its target stars within that field.

As described in \citet{jenkins2010overview}, the Kepler Science Operation Center (KSOC) pipeline processed long cadence data performing the following steps: the raw data are calibrated at the pixel levels (removal of CCD bias voltage and cosmic rays, flat field corrections, etc.), and sky-background subtracted, hence a simple aperture photometry (SAP) is performed for each target star. The optimal aperture is determined automatically to maximize the S/N for the given star \citep{2010ApJ...713L..97B}. This process results in the extraction of the stellar flux as a function of time, indicated as the SAP light curve, which is then corrected from systematics and other errors \cite[Presearch Data Conditioning (PDC) SAP light curve]{2010SPIE.7740E..1UT} and searched for events by the Transiting Planet Search (TPS) module \citep{2010SPIE.7740E..0DJ}. 

Once a TCE is detected, the light curve is searched again for additional events, after all occurrences of the first one are masked. This is done iteratively, until up to ten TCEs are identified. For an extensive description about the procedure of transits detection the reader may refers to \citet{2002ApJ...564..495J, 2010SPIE.7740E..0DJ} and \citet{christiansenkepler}.
For all TCEs, a set of supplementary statistics are computed by the Data Validation (DV) module, which can aid the vetters in assigning dispositions to them \citep[see e.g.][]{2016ApJ...822...86M,2016PASP..128g4502M,Coughlin_2016,2018ApJS..235...38T}). 

All the {\it Kepler} data used in this paper can be found in MAST \citep{10.17909/T98304}\footnote{\url{http://archive.stsci.edu/missions/kepler/lightcurves/tarfiles/DOI_LINKS/Q0-17_LC+SC/}}.

\subsection{TESS Data} 
\label{sec:tess_data}
TESS was designed to perform photometric observations for $\sim$200,000 target stars at 2-min cadence (short cadence data) and simultaneously collect images of its entire field of view every 30-min (Full-Frame Images, long cadence data). The sampling rate of Full-Frame Images was reduced from 30 to 10 minutes starting from the first TESS Extended Mission (July 2020). 
The features related to each observed star are stored in the TESS Input Catalog (TIC) \citep{2019AJ....158..138S}. This catalog is periodically updated and its information (e.g. stellar properties, ancillary data) are computed by using Gaia DR2 \citep{2018A&A...616A..10G} as a base catalog and other photometric catalogs like APASS \citep{2018ApJ...867..105T}, UCAC4 and UCAC5 \citep{2013AJ....145...44Z,2017AJ....153..166Z}.
The data for the target stars are reduced and analyzed by the Science Processing Operations Center (SPOC) pipeline \citep{jenkins2016tess}, which is strongly based on the KSOC pipeline described above, with some modifications (e.g. removal of spurious signals due to the absence of shutters on the TESS cameras). 
Calibrated images are computed both for short and long cadence data.
The optimal aperture for target stars sampled at 2-min is determined as in KSOC so that the SAP light curves \citep{2010SPIE.7740E..1UT,2020ksci.rept....6M} are extracted. The SPOC PDC \citep{2012PASP..124.1000S} module applies a set of corrections to the SAP light curve (e.g. rejection of outliers and instrumental systematics). This PDCSAP light curve is searched for TCEs by the TPS module as in KSOC. 
The DV module computes a set of supplementary features for each TCE identified by TPS, publicly available on the Mikulski Archive for Space Telescope (MAST)\footnote{\url{https://exoplanetarchive.ipac.caltech.edu}.}. 
Full-Frame Images are processed by the MIT's Quick Look Pipeline (QLP) that extracts light curves for all stars with TESS magnitude $\leq$ 13.5 \citep{2020RNAAS...4..204H,2020RNAAS...4..206H,2021RNAAS...5..234K,2022RNAAS...6..236K} \footnote{QLP data can be found at \url{https://archive.stsci.edu/hlsp/qlp} \citep{10.17909/t9-r086-e880}.}. In accordance with the methodology employed by SPOC, QLP generates normalized light curves from the optimal aperture (SAP light curves). This pipeline removes stellar variability from SAP light curves by fitting a basis spline (B-spline) with spacing in the range [0.3, 1.5] \citep{2014PASP..126..948V}. This process is repeated iteratively until convergence. 
At each iteration, $3\sigma$ outliers are identified and discarded before re-fitting the B-spline. The QLP searches the detrended light curves (KSPSAP light curves) for TCEs using the Box Least Squares algorithm \citep[BLS;][]{2002A&A...391..369K}.
In our work, we use the normalized, detrended light curve generated by all these pipelines, phase-folded over the period provided by the TCEs and binned. We show in section \ref{sec:test} that our results are independent of the pipeline used to extract the light-curves.
Hence, for a planetary transit detected from a generic survey to be vetted by our model, we just need to have the standardized photometric data, the epoch of the observed transit-like signal and the period of the candidate planet. 

All the {\it TESS} data used in this paper can be found in MAST \citep{10.17909/t9-nmc8-f686}\footnote{\url{https://archive.stsci.edu/tess/bulk_downloads/bulk_downloads_ffi-tp-lc-dv.html\#lc}}.

\section{Catalogs of currently available labelled TCEs}
\label{sec:tce_and_labels}
To achieve our goal of developing a simple and adaptable automatic triage model, we demonstrate its flexibility by evaluating it on various datasets, including Kepler and TESS TCEs, used both individually and in combination. Details of the datasets used are provided below.

\subsection{Kepler dataset}
\label{sec:catalog_kepler}
The Kepler dataset we used is the Autovetter Planet Candidate Catalog \citep{catanzarite2015autovetter} for Q1-Q17 \citep{2015ApJS..217...18S} Data Release 24 (DR24) and Data Release 25 (DR25, \citet{2018ApJS..235...38T}).
\begin{itemize}
    \item The DR24 catalog consists of 20,367 TCEs detected by the KSOC pipeline in Kepler PDCSAP light curves. These TCEs have been automatically labelled by Autovetter \citep{catanzarite2015autovetter}. To enhance the accuracy of the training set labels, we followed the approach proposed by SV18 and removed all TCEs classified as unknown. In order to binarize the labels, we associated 3,600 TCEs planet candidates to the PC class and the remaining 12,137 TCEs, including 9,596 astrophysical false positives and 2,541 non transiting phenomenon, to the NPC class. 

    \item Kepler Q1-Q17 DR25 contains 34,032 transit signals detected in the 17 quarters of Kepler primary mission dataset. The detection process of these TCEs is described in \citet{2016AJ....152..158T}. To generate our set of PCs and NPCs TCEs from this catalog, we performed the following steps. First, we removed all TCEs with rogue flag set to 1. These are TCEs with less than three detected transits, added to the DR25 catalog because of  a bug in the Kepler pipeline. For our PC class, we considered the set of 2,726 confirmed and 1,382 candidate planets from the Cumulative KOI catalog. This catalog was created with the intent of providing in one place the most accurate information about all the Kepler TCEs dispositioned as confirmed/candidate planet or false positive. We included into our NPC class the 3,946 TCEs labelled as false positive in the Cumulative KOI table and the 21,098 TCEs from the DR25 catalog that are not in the Cumulative KOI table.
\end{itemize}

\subsection{TESS dataset}
\label{sec:catalog_tess}
We used four catalogs of TESS TCEs: the Exoplanet Follow-up Observation Program (ExoFOP), TESS Triple 9 and the catalogs provided by YU19 and TEY23.

\begin{itemize}
    \item The catalog from the Exoplanet Follow-up Observation Program (ExoFOP)\footnote{We downloaded the catalog from \url{https://exofop.ipac.caltech.edu/tess/view_toi.php} on January 12, 2023.}. From this catalog, we discarded 1,453 TCEs overlapping with the dataset YU19. We also removed $\sim$560 TCEs for which there is no indication of the detection pipeline, i.e., SPOC or QLP. The remaining set of TCEs, which we define as $\mathcal{D}_E$, contains 3,133 planets (including known and confirmed planets, planet candidates, ambiguous planetary candidates and nearby planet candidates\footnote{Nearby planet candidates are TCEs consistent with a planet candidate signature but their detected transit occurs around a nearby star.} that we labelled as PC) and 419 false positives (including false alarms and eclipsing binaries, that we labelled as NPC) with dispositions assigned by the TESS Follow-up Observing Program Working Group (TFOPWG) \citep[][]{2019AAS...23314009A}.
    
    \item The TESS Triple 9 catalog ($\mathcal{D}_{TT9}$, \citet{2022MNRAS.513..102C, 2023MNRAS.521.3749M}). This catalog has been compiled within the framework of the NASA Citizen Science project Planet Patrol\footnote{\url{https://www.zooniverse.org/projects/marckuchner/planet-patrol/}}, an initiative engaging both scientists and amateur astronomers in the vetting process to expedite the analysis of a large number of TCEs within a short time frame. Initially, this catalog contained 999 TCEs for which \citet{2022MNRAS.513..102C} provided dispositions. A new set of 999 TCEs has been recently labelled and added to the catalog by \citet{2023MNRAS.521.3749M}. In both cases, the authors determined the TCEs dispositions by manually examining the outputs produced by \texttt{DAVE} \citep{2019AJ....157..124K} for each TCE. The samples of this catalog have been labelled as candidate planet, false positive and potential false positive. The latter label is used when human vetters are not fully convinced of the non-planetary nature of the given TCE . In the process of binarizing these labels, we decided to discard all potential false positives (amounting to 251 TCEs) to reduce the level of uncertainty affecting our test set. By further removing 187 TCEs in common with YU19, we obtain a final catalog of 1,314 PCs (including candidate planets) and 276 NPCs (including false positives) homogeneously vetted.
    
    \item The catalog of YU19 ($\mathcal{D}_{YU19}$) contains $\sim$16,700 TCEs detected in QLP SAP light curves from TESS sectors [1-5]. In total, there are 505 events labelled as planet candidates, 2,247 as eclipsing binaries, 976 as V-shape, 816 as instrumental systematics and 12,170 events labelled as junk. The authors have manually ascribed these dispositions through the implementation of a predefined set of rules, aiming to guarantee uniformity in the labelling process (see Section 2.2 of their paper for further details). During the binarization step, we consider as PCs the 505 TCEs they classify as planet candidates and all other 16,209 TCEs as NPCs.
    \item The catalog provided by TEY23 ($\mathcal{D}_{TEY23}$), consisting of a subset of TCEs detected by QLP in TESS long cadence data that the authors have manually vetted. This subset is composed of 24,952 TCEs (including 8,992 TCEs detected in Sector 13; 13,372 TCEs detected from Sector 14-26; 2,588 TCEs from Sectors 27-39). The authors classified these TCEs as: ``periodic eclipsing signal'', ``single transit'', ``contact eclipsing binaries'', ``junk'', ``not-sure'' (see Section 2.4 of their paper for further details on the labeling process). By discarding (i) 5,340 TCEs for which the authors did not provide a consensus label and (ii) all the TCEs labelled as ``single transit'' and ``not-sure'',  we obtained 2,613 periodic eclipsing signals, which include planet candidates and non-contact eclipsing binaries, and 16,5290 not-planet, including 738 ``contact eclipsing binaries'' and 15,791 ``junk''. 
\end{itemize}

Because the TESS catalogs exhibit an imbalance toward one of the two classes, we generated a single catalog of TESS TCEs by merging data from the $\mathcal{D}_E$, $\mathcal{D}_{TT9}$, $\mathcal{D}_{YU19}$ datasets. We then obtained a more balanced catalog of TESS ($\mathcal{D}_T$) TCEs from which we could derive significantly more reliable model evaluation metrics. In the discussion of the results obtained on this test set (Section \ref{sec:test_tess}), we have beforehand discarded 355 samples which disposition differs between $\mathcal{D}_E$ and $\mathcal{D}_{TT9}$ catalogs. 
To evaluate the predictive capabilities of the model when trained-tested on cross-mission data, we also generated a catalog containing TCEs from $\mathcal{D}_T$ and $\mathcal{D}_{K24}$.
We discuss the motivation for creating this dataset in  Section \ref{sec:crossmission-data} . 

Table \ref{tab:tce_catalogs} summarizes the details of the different catalogs presented in this section. 

\begin{table}\centering
    \resizebox{\textwidth}{!}{
    \hspace{-2cm}\begin{tabular}{llllllll}
    \hline 
    \textbf{Dataset} & \textbf{Name} & \textbf{Positive class} & \textbf{Negative class} & \textbf{Imbalance ratio (\%)} & \textbf{Task} & \textbf{Reference} \\
    \hline 
        Kepler Q1-Q17 DR24 & $\mathcal{D}_{K24}$ & 7,200~\footnote{\label{note1}The number of positive training samples was doubled with data augmentation techniques, as described in Section \ref{sec:data_prep}.} & 12,136 & 37~:~63 & binary & (1)\\
        Kepler Q1-Q17 DR25 & $\mathcal{D}_{K25}$ & 8,216~\textsuperscript{\ref{note1}} & 24,657 & 24~:~76 & binary & (2)\\
        TESS ExoFOP & $\mathcal{D}_{E}$ & 2,976 & 576 & 84~:~16 & binary & (3)\\
        TESS Triple 9 & $\mathcal{D}_{TT9}$ & 1,314 & 527 & 71~:~29 & binary & (4)\\
        TESS YU19 & $\mathcal{D}_{YU19}$ & 996~\textsuperscript{\ref{note1}} & 16,154 & 6~:~94 & binary & (5)\\
        TESS TEY23 & $\mathcal{D}_{TEY23}$ & 4,830(E) & 700(B);12,318(J) & 4~:~27~:~69 & multi-class & (6)\\ \hline
        \textbf{Dataset combination} & & & & & & \textbf{Source} \\
        TESS (total) & $\mathcal{D}_T$ & 5,286 & 17,257 & 23~:~77 & binary & $\mathcal{D}_{E} \cup \mathcal{D}_{TT9} \cup \mathcal{D}_{YU19}$\\
        TESS + Kepler DR24 & $\mathcal{D}_{TK}$ & 12,486 & 29,393 & 30~:~70 & binary & TESS $\cup~\mathcal{D}_{K24}$ \\
        \hline 
    \end{tabular}
    }
    \caption{Composition of the different TCEs datasets used in this work. For each dataset, we provide: its name we use to refer to them in this work; the distribution of samples and imbalance ratio between the two classes; the type of task (binary or multi-class classification) for which they were used to train and test the model. The combined datasets are also listed, along with their respective class distributions and catalog sources. References: (1) \citet{catanzarite2015autovetter}; (2) \citet{2018ApJS..235...38T}; (3) \citet{2019AAS...23314009A}; (4) \citet{2022MNRAS.513..102C, 2023MNRAS.521.3749M}; (5) \citet{2019AJ....158...25Y}; (6) \citet{2023AJ....165...95T};}
    \label{tab:tce_catalogs}
\end{table}

\section{Standardization of input data}
\label{sec:data_prep}
Here we describe how we standardized the light curves for each TCE in the datasets listed in Table \ref{tab:tce_catalogs}. 

We represent each TCE as a phase-folded light curve divided into 201 bins. We first create a multi-sector light curve by stitching light curve segments from the different \textit{Kepler} quarters or TESS sectors. 
For \textit{Kepler} TCEs, we concatenate all the related PDCSAP light curves as in SV18. 
For TESS TCEs, we only stitch the SAP light curve segments in the same year in which the TCE was discovered. For example, if an event was discovered during TESS observation year 2, we use the light curves segments from TESS sectors [14,26] to generate the final light curve. This is done to prevent that the uncertainty on the transit parameters propagate onto distant observations leading in a mismatch of the folded data, as shown in Figure \ref{fig:tic_transitmask_shift}.

Then, we remove all outliers (flux measurements above $\pm \,3\sigma$ from the median flux) from the stitched light curve \citep{fiscale2021gpu} and any long-term variability in the signal that is not related to the TCE. 

\begin{figure}
    \centering
    \includegraphics[width=0.9\textwidth]{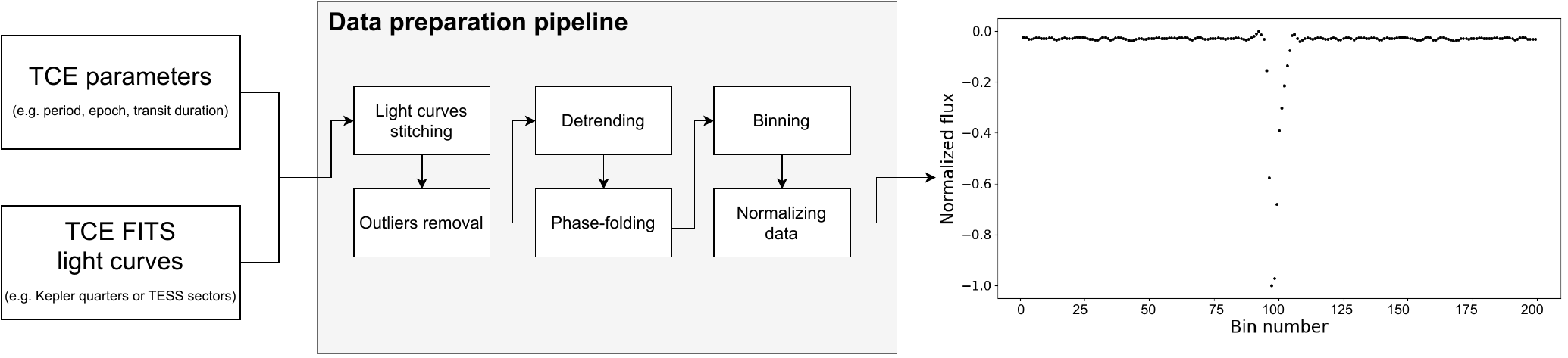}
    \caption{Data flow of our preparation pipeline. For each TCE, we downloaded the related FITS light curve files from MAST. These light curves are concatenated in a single signal that is cleaned from not-a-numbers and outliers. The variability in the observed flux not related to the TCE's transits is flattened in the detrending module. The resulting signal is folded over the TCE's period. We finally binned and normalized the signal in order to produce a 201-length global view with median flux and transit depth set to 0 and -1, respectively.}
    \label{fig:data_prep}
\end{figure}

\begin{figure}[h]
    \centering
    \includegraphics[width=\textwidth]{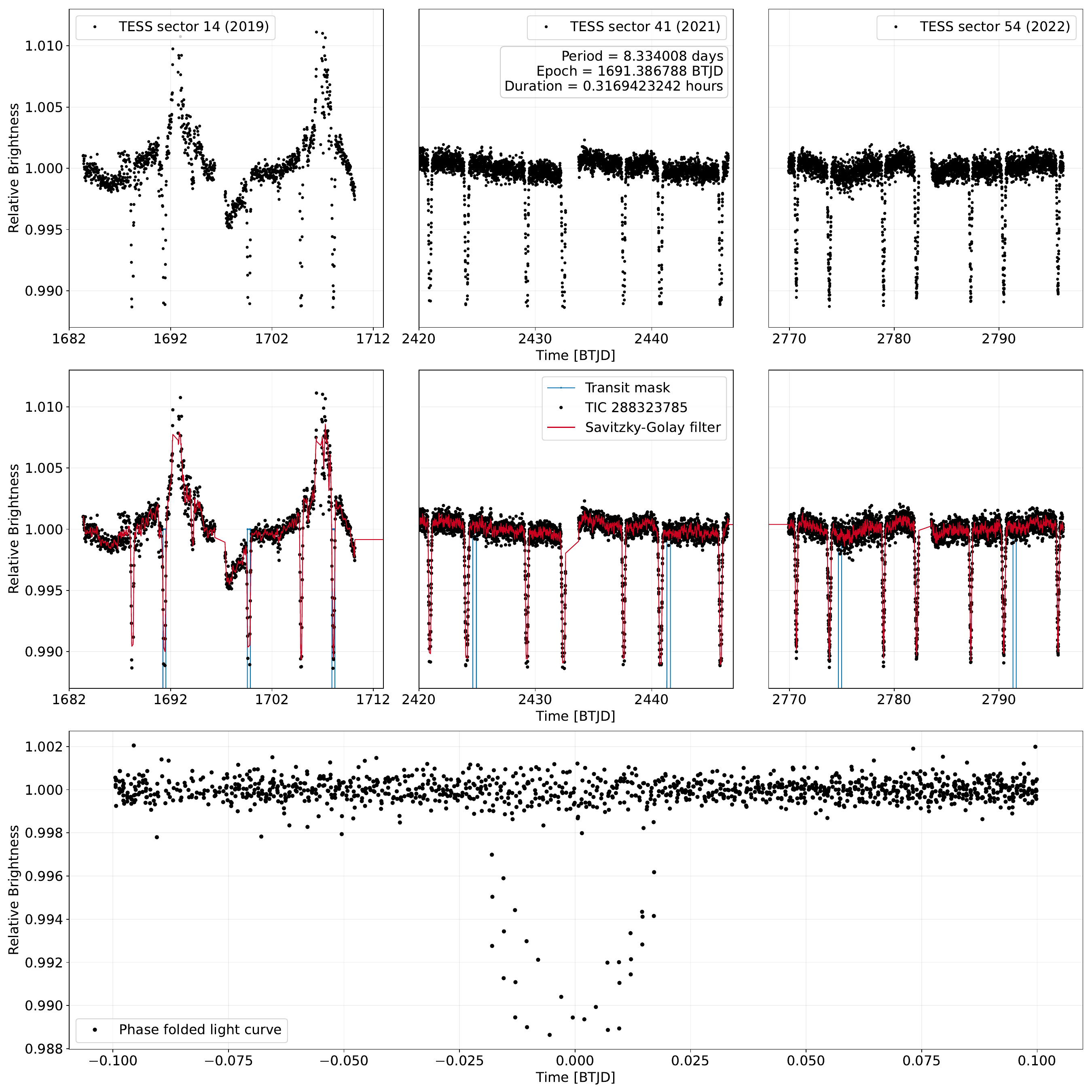}
    \caption{An example of incorrect generation of the global view. For TIC 288323785, TESS acquired photometric data over a span of 3 years (top panel). The transit mask (blue solid line), generated by using the period, epoch, and duration from the TCE, fits well the data from sector 14 (middle-left panel) but not the later ones (second and third middle panels). When flattening the light curves (black dots) dividing by the interpolating polynomial (red dashed line), transits from sectors 41 and 54 are not preserved. As a result, the light curve is not correctly phase-folded (bottom panel).}
    \label{fig:tic_transitmask_shift}
\end{figure}

We perform this by dividing flux values over the interpolating polynomial computed using the Savitzky-Golay filter \citep{10090063}. 
The flux values corresponding to the TCE transits are masked during this step so that they are preserved. This detrending process aligns with recent data-driven approaches proposed for astrophysical applications \citep{orsini2025data}.
Our detrended signal is phase-folded on the relative period and binned with a time bin size of 30min.
As in YU19, we linearly interpolate over any void bin to generate a uniform input signal of length 201. Following SV18, the final step of our data preparation pipeline consists of normalizing the binned flux so that the median value is zero and the maximum transit depth is fixed at -1. By implementing this data normalization, we enhanced the numerical stability of the model; however, the information on the transit depth is lost. The binned and normalized light curve is called the \textit{global view} and it represents the input of our model. The pre-processing pipeline described here is schematized in Figure \ref{fig:data_prep}.

\begin{figure}
    \centering
    \includegraphics[width=\linewidth]{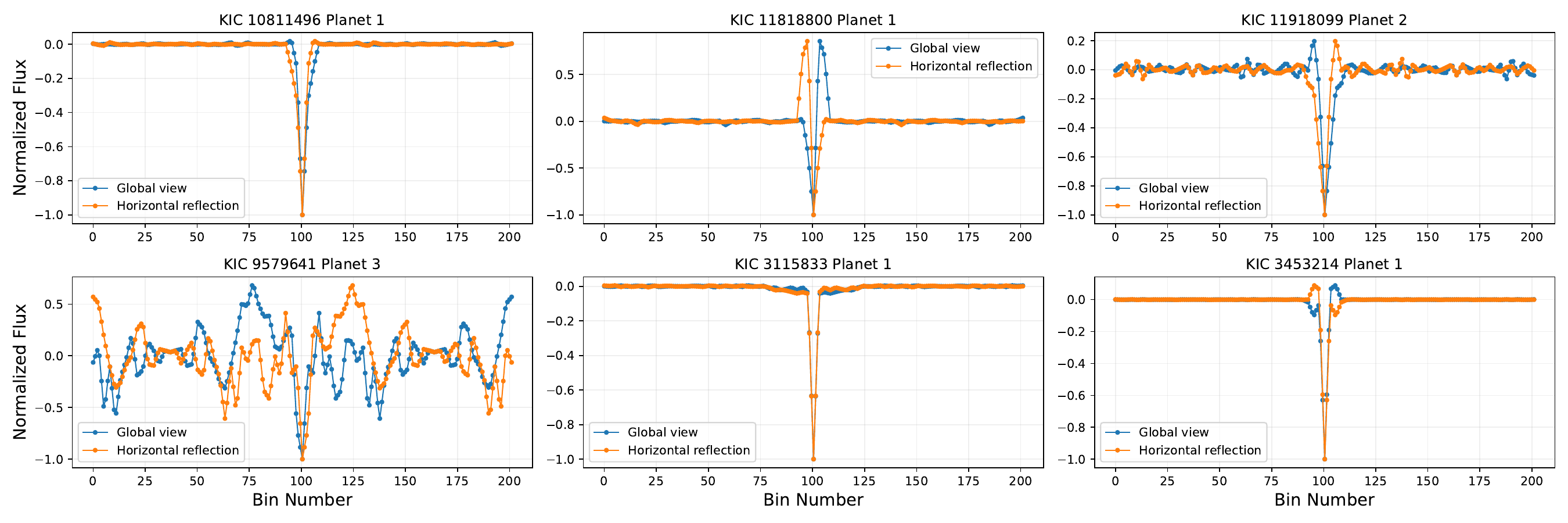}
    \caption{An example of the output obtained by the horizontal reflection. From top left to bottom right we display six global views (blue lines) generated for six planet candidates KOIs from Kepler Q1-Q17 DR25. For each global view we provide its horizontally reflected version (orange lines). The horizontal reflection procedure is applied over all the datasets highly imbalanced toward the class of NPC.}
    \label{fig:horizontal_reflection}
\end{figure}

In order to perform the data augmentation, we also produce a horizontal reflection of the PCs global views to obtain additional samples (Figure \ref{fig:horizontal_reflection}).
We applied this strategy specifically to the unbalanced dataset in Table 1, where the number of negative samples far exceeds that of positive ones, to mitigate the imbalance between positive (PC) and negative (NPC) TCEs. 

Due to the significant computational cost required by this pipeline, we employed parallel distributed strategies to optimize computational time \citep{de2019distributed}. 

\section{Network architecture}
\label{sec:model_architecture}
The architecture of our model is a thinned version of the one proposed by \citet{2014arXiv1409.1556S}, which is a one-dimensional CNN. This kind of network belongs to the class of Deep Learning models and have proved their effectiveness in processing time-series data in different research fields \citep[see e. g.][]{zhao2017convolutional, durairaj2022convolutional}.
A CNN is composed by two main branches: (i) a \textit{feature extraction} branch - in which ever more complex features and non-linear relationships (defined as feature maps) are extracted from input data; (ii) a \textit{classification branch} - where the output of the previous branch is processed feed-forwardly through at least one layer of neurons. Typically, the sigmoid activation function is used to map the output of the classification branch in the range [0,1].

Our model takes the global view as input and processes it through the feature extraction branch. This branch is composed of five convolutional blocks. Each block has one convolutional layer with filter size 3 and a number of convolutional filters increasing by power of two in the range [16, 256]. 
During the convolutions, we preserve the length of the input feature maps by using the padding technique. Otherwise, the length of the feature maps generated by the convolutional blocks would progressively diminish to 0, resulting in a complete information loss. We then apply the Rectified Linear Unit (ReLU) activation function to the output provided by the convolutional layer. At this stage of processing, we prevent feature maps of the model from overfitting by using the spatial dropout with dropout probability rate $\rho$ fixed to 0.2. The convolutional block structure ends with a max-pooling layer and a batch normalization layer \citep{ioffe2015batch}. The max-pooling layer reports the maximum output within a 5-length neighborhood. The batch normalization layer regularizes the output of the block by introducing both additive and multiplicative noise.

The feature tensor produced by the feature extraction branch is flattened and delivered to the classification branch. This branch is composed of a single dense block containing a fully connected layer of 512 neurons and a dropout layer. 

We used a single fully-connected layer because it is adequate to guarantee the best approximation property \citep{cybenko1989approximation,hornik1989multilayer}. 
\citet{2022A&C....4100654V} showed that by reducing the number of model's parameters (i.e. the model complexity) its performance are not subject to any deterioration. Typically, the model complexity mainly depends on the number of the fully-connected layers composing the classification branch. A network with a single fully-connected layer of $n$ neurons shows $n^2$ connections (i.e. parameters). By increasing the number of fully-connected layers to $H$, the number of parameters to be optimized becomes $n^2 \times (H-1)$. To understand this, consider that $n \times n$ is the number of connections between two consecutive layers, and with $H$ layers the number of connections between them is $H-1$. Typically, the density of connections in a fully-connected network grows quadratically with the number of $n$ neurons. The higher the model complexity, the larger the training set size required. Since for labelled TCEs we cannot rely on such a large training set yet, we have to reduce the model complexity to prevent the model from overfitting or underfitting.

\begin{figure}
    \centering
    \includegraphics[width=0.5\textwidth]{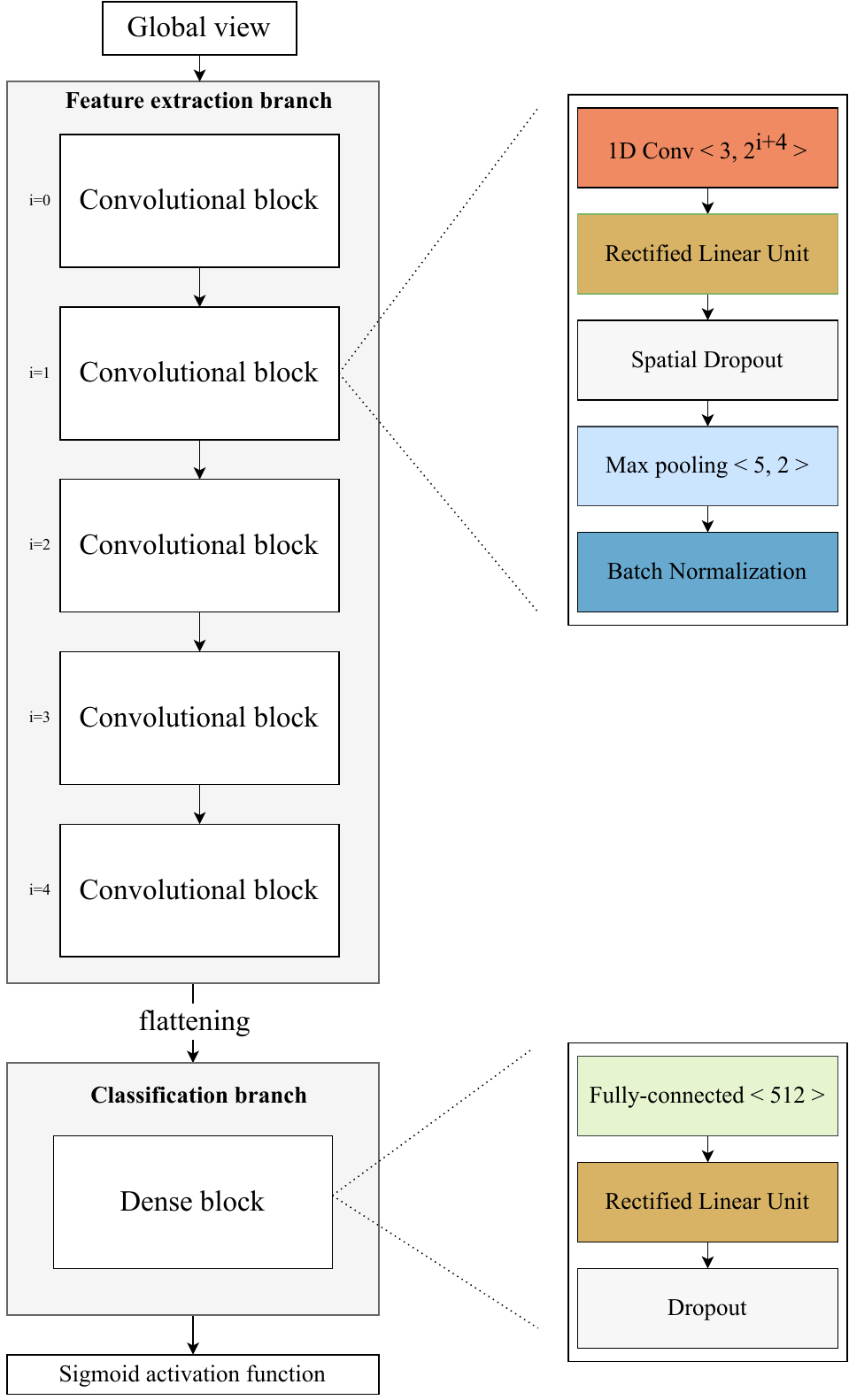}
    \caption{The architecture of the CNN model.
    The feature extraction branch is responsible for features extraction from the global view. The feature tensor produced by the last convolutional block is subjected to flattening. Then, classification is executed through utilization of a multi-layer perceptron within the classification branch.}
    \label{fig:model_architecture}
\end{figure}

The output layer of our model is a single neuron provided with the sigmoid activation function. We binarize the value produced by this function through a thresholding process. For output values greater than 0.5, the input TCE is labelled as PC. Otherwise, it is labelled as NPC. Figure \ref{fig:model_architecture} depicts the architecture of our model.

The source code necessary for the deployment and training of the CNN architecture, as well as that essential for the generation of input samples are available on GitHub \footnote{\url{https://github.com/stefanofisc/dartvetter_apj}}.

\subsection{Combining training data from different space missions}
\label{sec:crossmission-data}

In ML, exploiting data from different domains can help the model to reduce overfitting to specific datasets. In our case, training the model on \textit{Kepler} and TESS TCEs brings the following benefits: $i)$ it allow us to increase the number of TCEs available for training; $ii)$ it allows us to improve the network generalization performance.
Hence, using TCEs from both \textit{Kepler} and TESS datasets allows us to train a more robust and versatile model, improving its performance in the detection of exoplanets across different surveys.
In \citet{fiscale_wirn}, we optimized the parameters of the model by combining \textit{Kepler} and TESS TCEs using Transfer Learning \citep{ribani2019survey} and observed that the optimal performance (F1-score of 84\%) was achieved by pre-training the model on \textit{Kepler} TCEs and subsequently fine-tuning on TESS TCEs. However, during the process to further simplify the model's architecture, we found that \texttt{DART-Vetter} performs better when trained on the entire set of TCEs at once.

\subsection{Model selection}
\label{sec:model_selection}

The network architecture of \texttt{DART-Vetter} was determined through a careful and extensive process of \textit{model selection}, aimed at determining the optimal model in minimizing the fraction of misclassified planetary signals.

First, we defined a base configuration consisting of a neural network with fixed hyperparameters. We undertook a grid search process to train and test a set of configurations, by varying the number of convolutional blocks, dense block structure and the values of the hyperparameters such as learning rate, batch size and dropout probability rate. All these configurations shared the same training and test sets. Each configuration was evaluated based on its ability to balance model complexity and predictive capabilities. This evaluation allowed us to identify the best model able to generalize on signals from different transiting surveys. We provide further details on the model selection phase in the Appendix \ref{sec:appendix_a}. The resulting model, named \texttt{DART-Vetter}, combines a compact structure with high predictive performance, making it a robust tool for the classification of TCEs.

\subsection{Training details}
\label{sec:model_training}
Once we fixed the hyperparameters, we initialize the parameters of our network drawing values from a uniform distribution \citep{klambauer2017self}. These parameters are optimized by using the Adam's optimization algorithm \citep{kingma2014adam} to minimize the loss function. The learning rate value used at this stage is $\alpha$=$10^{-3}$ and the batch size is 128. As loss function, we employ the binary cross-entropy, that measures the difference between the real dispositions and the ones predicted by the model. 
 As highlighted in the "Imbalance Ratio" column of Table \ref{tab:tce_catalogs}, all our datasets exhibit an imbalance towards one of the two classes. To address class imbalance, we use the class weighting technique. More precisely, to differently weigh the model's predictive error on positive and negative samples we employed the inverse class frequency method \citep{lertnattee2004analysis}.

We make use of the dropout technique, which allows us to perform the training only once:
given a network with $n$ neurons, or units, the dropout allows to train $2^n$ different sub-networks 
\footnote{The feasibility of training all $2^n$ networks relies on the condition that this quantity enables the generation and training of all possible sub-networks within a reasonable amount of time.}. 
During the training, each sub-network is obtained by dropping some units, with a probability $\rho$, from the entire network architecture, but as all sub-networks share the same parameters, we save a significant amount of memory \citep{lecun2015deep} with respect to other networks in which the training is done with standard model averaging \citep{claeskens2008model}. 
In that case, it is common to use ensembles of $K$ = 5 to 10 neural networks \citep{szegedy2015going}, each of them with their own set of parameters. Because of the independence of these parameters between any two different networks of the ensemble, all $K$ models have to be trained separately. 
In the context of exoplanets detection, this technique has been widely adopted starting by SV18. However, training multiple independent models is highly time consuming, particularly for large architectures such as \texttt{ExoMiner} and \texttt{AstroNet-Triage-v2}. These models rely on \textit{K}-fold cross-validation, requiring \textit{K} separate training and evaluation runs, with the final TCEs prediction achieved by averaging the \textit{K} independent classifications. This process inevitably complicates the reproducibility of these models.

For these reasons, the use of model averaging can be very expensive in terms of computational complexity, whilst dropout allows to train many more than $K$ networks but in the same time required to train a single network, i.e. $\sim$27 seconds in our case. 
During the model selection, we tested values for $\rho$ in the range [0.2, 0.5] as suggested in \citet{srivastava} and found the best value of $\rho=0.2$.

\section{Experimental results}
\label{sec:test}
In this section we present the experimental results, assessing the network performance across various datasets, tasks and evaluation metrics. The results for binary classification task are detailed in Section \ref{sec:test_binary}, while Section \ref{sec:test_multi} focuses on the performance of \texttt{DART-Vetter} in the context of multi-class classification.  

We use the conventional metrics in the field of machine learning in order to evaluate the model, namely:
\begin{itemize}
    \item \textit{Accuracy} --  the ratio of correctly predicted TCEs to the total number of samples in the dataset:
    \begin{equation*}
        \label{eq:accuracy}
         \frac{\text{true positives + true negatives}}{\text{true positives + true negatives + false positives + false negatives}},
    \end{equation*}
    where true positives (TP) and true negatives (TN) are TCEs correctly identified as positive and negative by the classifier, respectively; false positives (FP) are TCEs identified as positive but labelled negative, and false negatives (FN) are the samples identified as negative but labelled positive;

    \item \textit{Misclassification rate} -- the fraction of TCEs inaccurately classified by the model, computed as:
    \begin{equation*}
        \label{eq:misclassification_rate}
        \text{Misclassification rate}: \frac{\text{\# incorrect predictions}}{\text{\# total predictions}} = \frac{\text{false positives + false negatives}}{\text{\# total}},
    \end{equation*}
    
    where total is  the total number of samples of a given dataset, i.e. the sum of TP, FN, FP and FN;
    
    \item \textit{Precision} -- 
    fraction of planets correctly classified by the model among all TCEs that the model has classified as PCs:

    \begin{equation*}
        \label{eq:precision}
        \text{Precision}: \frac{\text{true positives}}{\text{true positives + false positives}}
    \end{equation*}

    \item \textit{Recall} -- fraction of planets correctly classified by the model:
    \begin{equation*}
        \label{eq:recall}
        \text{Recall}: \frac{\text{true positives}}{\text{true positives + false negatives}}
    \end{equation*}
    
    \item \textit{F1-score} -- precision and recall are averaged to obtain the \textit{F1-score}, which is often used  to assess the overall effectiveness of a classifier's performance: 
    \begin{equation*}
        \text{F1-score}: 2 \cdot \frac{\text{Precision} \cdot \text{Recall}}{\text{Precision} + \text{Recall}}
    \end{equation*}

\end{itemize}

\begin{figure}
    \centering
    \includegraphics[width=\textwidth]{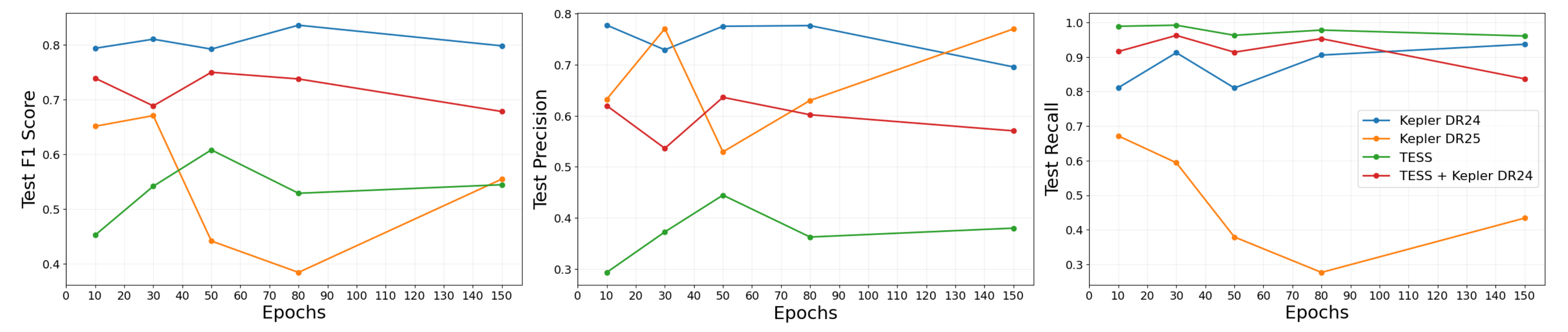}
    \caption{Variation of the main evaluation metrics (F1, precision and recall) as the number of training epochs increases. These values have been computed on the test set of each dataset. For sake of clarity, concerning TESS data we display the performance of \texttt{DART-Vetter} on the dataset TESS that includes ExoFOP, Triple 9 and YU19.}
    \label{fig:performance_over_epochs}
\end{figure}

\subsection{Binary classification on single/cross domain data}
\label{sec:test_binary}
We report the performance of the model in solving the task of binary classification on TCEs belonging either to a specific and multiple surveys. All the tests presented in the subsections below were conducted following this methodology:

\begin{itemize}
    \item [] STEP 1: We fixed a dataset from Table \ref{tab:tce_catalogs}, a training-test set split of 80-20 and a grid of training epochs of [10, 30, 50, 80, 150];

    \item [] STEP 2: For each value of the grid, \texttt{DART-Vetter} was trained and evaluated. At the end of each training-test phase, we computed the scores for all the metrics. This procedure returned 10 values for each evaluation metric (5 corresponding to the training set and 5 to the test set);

    \item [] STEP 3: We analyzed the variation across these 5 values to assess the model's convergence speed in solving the task at hand on the fixed dataset. To conduct this analysis, we computed the mean and standard deviation for each group of 5 values. For a given metric, high standard deviation values indicate that the model struggles to converge toward optimal performance. We illustrate the average and variation of values for precision, recall and F1-score computed on each dataset in Figure \ref{fig:performance_average_errorbars}.

\end{itemize}

The confusion matrices computed on all the test sets are reported in Table \ref{tab:confmat}. Additionally, in Appendix \ref{sec:appendix_b} we examined the variation of model performance when training and testing on different dataset splits. The training-test splits considered in this analysis were 60-40, 70-30, 80-20 and 90-10. These results are displayed in Figure \ref{fig:performance_over_datasplit}.

Here, we want to remark that we provide a complete set of evaluation metrics and explicitly detailing their computation as we strive for transparency in our work. Our approach aims to provide a comprehensive and well-documented evaluation framework, which ensures greater clarity and facilitates reproducibility.

\subsubsection{Test 1. Application on Kepler Q1-Q17 Data Release 24}
The first set of TCEs on which we evaluated the predictive capabilities of the model is $\mathcal{D}_{K24}$. The results we obtained at the end of the three outlined steps are detailed below.

The model achieves optimal performance when trained for 80 epochs, obtaining a precision of 0.77, recall of 0.90 and F1-score of 0.83. 
The values of the metrics remain stable as training epochs vary on both training and test sets (blue lines in Figure \ref{fig:performance_over_epochs}). On the test set, precision varies by 0.08 and recall by 0.12, with F1-score ranging of 0.04, reaching its maximum value for 80 training epochs. These small variations, visible with the relative error bars in Figure \ref{fig:performance_average_errorbars},  indicate a rapid convergence of the model to returning reliable predictions on this dataset.
The fractions of true negative and positive are 0.84 and 0.90, respectively. These rates are listed in the first row of Table \ref{tab:confmat}.

On this set of TCEs, \texttt{DART-Vetter} proves to be a robust model for performing triage due to the low false negative rate of 0.09 and misclassification rate of 0.13.   

\subsubsection{Test 2. Application on Kepler Q1-Q17 Data Release 25}
The Kepler Q1-Q17 Data Release 24 catalog does not include many long-period and low-S/N TCEs, which are instead included in the latest Kepler TCEs catalog: the Kepler Q1-Q17 Data Release 25. We evaluate \texttt{DART-Vetter} also on this catalog and detail the results below.

The best performance on the test set is obtained when the model is trained for 30 epochs, but the triage capabilities of DART-Vetter on this dataset are not reliable. The second row of Table \ref{tab:confmat} reports the confusion matrix calculated on the test set. The precision, recall and F1-score on the test set reach values of 0.77, 0.59 and 0.67, respectively. In fact, the model is not able to learn relevant information during training, obtaining a precision of 0.64 and a recall of 0.86, which are 26\% and 13\% lower than the respective values obtained during the training on Kepler DR24.

As the training epochs increase, there are significant changes in the values of F1-score, precision and recall on the test set (orange lines, Figure \ref{fig:performance_over_epochs}). As a result, the error bars calculated on these metrics and shown in Figure \ref{fig:performance_average_errorbars} are larger than those obtained on the Kepler DR24 test set. The wide variations observed in the metrics as the training epochs increase suggest that the model does not converge smoothly to a correct classification. Such fluctuations in the values are motivates the high false negative rate of 0.40.

\subsubsection{Test 3. Application on TESS data}
\label{sec:test_tess}
We now examine how \texttt{DART-Vetter} performs on TESS data. We initially trained and tested the model separately on ExoFOP, Triple 9 and YU19, following the steps 1, 2, and 3. As seen from Table \ref{tab:tce_catalogs}, all three catalogs are heavily unbalanced toward one of the two classes. Although we handled this issue by adopting class-weighting during training, such an imbalance still hampers the generalization capabilities of the model. For example, on test sets strongly unbalanced toward the PC class, \texttt{DART-Vetter} would obtain high recall and F1-score values simply by classifying all samples as PCs. For this reason, we present here the performance obtained by the model only on the TESS dataset (described in Table \ref{tab:tce_catalogs}), but the confusion matrices, accuracy and misclassification rate scores are listed in Table \ref{tab:confmat}. 

The best results on test set are achieved on 50 training epochs, with the model showing robust triage capabilities, as it achieves a recall rate of 0.96, that is higher with respect to any other dataset.

The green lines of Figure \ref{fig:performance_over_epochs} show a stability close to 1 in the recall on test set, as the number of training epochs increases. In contrast, the precision scores exhibit significant oscillations, indicating challenges in the model's ability to correctly classify TCEs belonging to the NPC class.
This difficulty is reflected on the high false positive rate of 0.34, as shown in the second to last row of Table \ref{tab:confmat}.

\subsubsection{Test 4. Application on cross-domain data}
ML models can achieve more robust predictive capabilities when trained on datasets from multiple domains. In order to do this, the data need to be standardized, ensuring invariance to the noise characterizing the original distributions from which the data are drawn. 
In this study, we perform this normalization by pre-processing the TCEs using the pipeline described in Section \ref{sec:data_prep}. The model was trained and tested on the $\mathcal{D}_{TK}$ dataset, following the methodology outlined in Steps 1, 2, and 3. The results obtained on this set of TCEs are presented below.

Optimal performance on $\mathcal{D}_{TK}$ are achieved after 50 training epochs, with precision and recall rates of 0.63 and 0.91 on the test set. The harmonic mean, that is the F1-score, is 0.75. 
The generalization capabilities of the model do not show significant variations along the training epochs (red lines, Figure \ref{fig:performance_over_epochs}). Moreover, the model achieves better performance than that obtained on the TESS dataset. Comparing the scores obtained between $\mathcal{D}_T$ and $\mathcal{D}_{TK}$, a great improvement is found in the overall values of F1-score and precision, while the recall value remains very high. The standard deviation calculated on the F1-score is reduced compared to that calculated on $\mathcal{D}_T$, indicating that the addition of TCEs from $\mathcal{D}_{K24}$ is useful to improve generalization.
The capabilities in performing a reliable triage on cross-mission data are confirmed by the low fraction of false negatives, corresponding to 0.07. The confusion matrix on test set is given in the last row of Table \ref{tab:confmat}.

Finally, we emphasize the benefit of training the model on the combined Kepler and TESS dataset. Using this dataset results in a significantly more reliable model, with recall stabilizing at approximately 92\%, while precision increases by $\sim$20\%. Additionally, as shown in Table \ref{tab:confmat}, the misclassification rate drops by 9\%, from 0.27 to 0.18, further demonstrating the effectiveness of this approach.

\begin{table}

    \begin{tabular}{llllllll}
    \hline 
        \textbf{Dataset} & \textbf{TN (\%)} & \textbf{FP (\%)} & \textbf{FN (\%)} & \textbf{TP (\%)} & \textbf{Accuracy} & \textbf{Misc. rate} \\ \hline 
        Kepler Q1-Q17 DR24 & 2056 (0.84) & 376 (0.15) & 136 (0.09) & 1304 (0.90) & 0.86 & 0.13 \\
        Kepler Q1-Q17 DR25 & 4645 (0.94) & 292 (0.05) & 669 (0.40) & 979 (0.59) & 0.85 & 0.14\\
        TESS ExoFOP & 11 (0.09) & 111 (0.90) & 8 (0.01) & 581 (0.98) & 0.83 & 0.16\\
        TESS Triple 9 & 16 (0.16) & 80 (0.83) & 15 (0.05) & 258 (0.94) & 0.74 & 0.25\\
        TESS YU19 & 2291 (0.71) & 923 (0.28) & 26 (0.12) & 190 (0.87) & 0.72 & 0.27\\
        TESS & 2250 (0.65) & 1196 (0.34) & 37 (0.03) & 957 (0.96) & 0.72 & 0.27\\
        TESS + Kepler DR24 & 4478 (0.77) & 1330 (0.22) & 192 (0.07) & 2307 (0.92) & 0.81 & 0.18\\ \hline 
    \end{tabular}
    \caption{Confusion matrices computed on all the test sets. For each dataset, we show the number of true negatives (TN), false positives (FP), false negatives (FN) and true positives (TP) along with the relative fraction (\%) to allow a fair comparison among all datasets. Accuracy and misclassification rates are also reported.}
    \label{tab:confmat}
\end{table}

\subsubsection{Comparison of network performance on different domains}
\begin{figure}
    \centering
    \includegraphics[width=\linewidth]{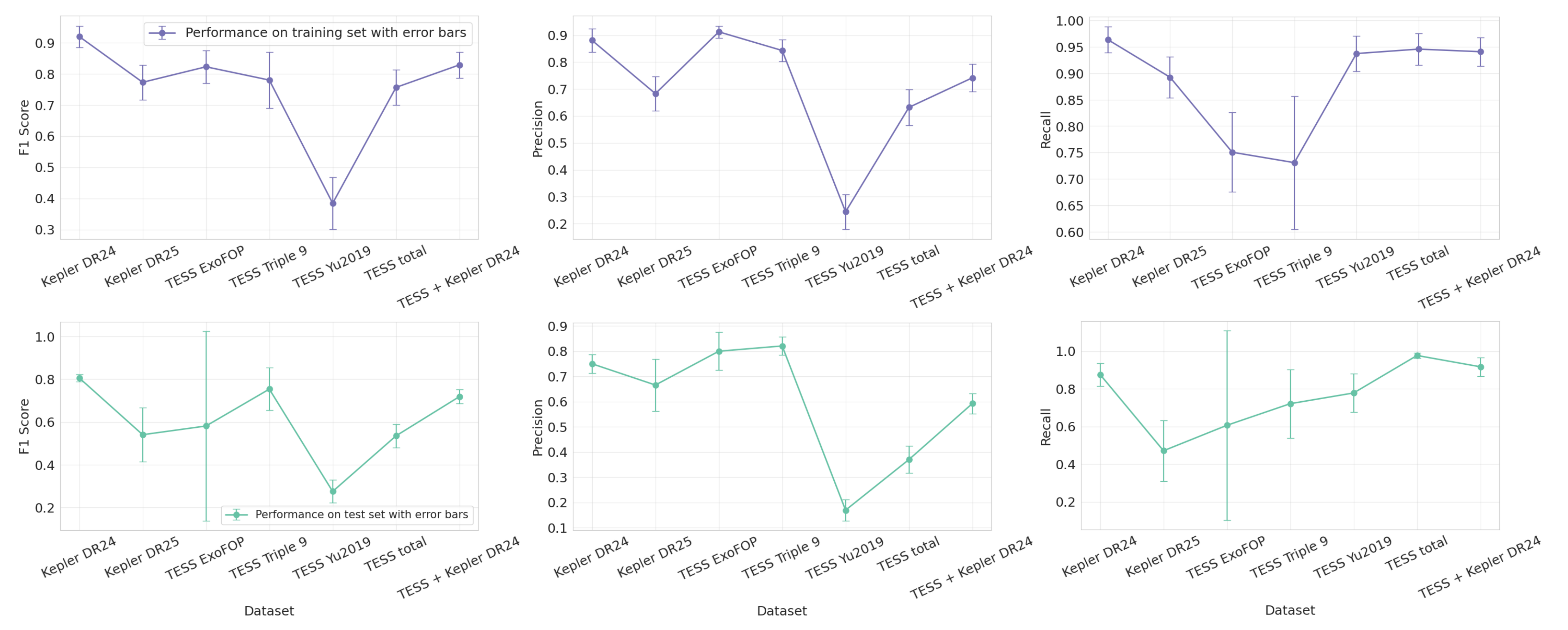}
    \caption{Mean values and error bars for F1-score, precision, and recall across all datasets. (Top panel, purple lines) Average scores and error bars computed for the specified evaluation metrics on the training sets. (Bottom panel, green lines) Corresponding values obtained on the test sets.}
    \label{fig:performance_average_errorbars}
\end{figure}

In this section we summarize and analyze the results from the above tests. Figure \ref{fig:performance_average_errorbars} shows the trend of the average values of F1, precision and recall as a function of the domain.

On the training set, the best performance is obtained on Kepler DR24, with precision, recall and F1-score close to 1. The error bar on the F1-score is smaller compared to the others, meaning that the model quickly converge. The performance worsens on Kepler DR25, on which \texttt{DART-Vetter} obtains an F1-score $\sim$13\% lower than on Kepler DR24. The wider error bars also indicate a slower convergence. This difference in performance on the two Kepler datasets reflects the difference on  the  S/N of TCEs in the two catalogs. 
The high precision values achieved on ExoFOP and TESS Triple 9 are related to the bias of the training set toward PC TCEs. In fact, the recall is relatively low and indicates that the model misses many PCs. 
More in-depth considerations of the true planetary nature of the PC TCEs are provided in Section \ref{sec:limitations_dispositions}. The opposite happens on the YU19 dataset, which is biased toward the NPC class. 
Finally, the performance of \texttt{DART-Vetter} on the entire TESS dataset is stable, with error bars decreasing and a second highest score on the training recall; which instead is the highest on the test set.
But the addition of Kepler DR24 TCEs to the TESS data produces even more robust performance: the model shows a faster convergence to optimal discrimination capabilities as discussed in the previous paragraph.

We do not discuss the corresponding results on the test sets (depicted by the green lines in Figure \ref{fig:performance_average_errorbars}) as they are in line with those presented on the training data.

\subsection{Multi-class classification on TESS data}
\label{sec:test_multi}
In order to compare the performance of our model with the ones available in literature, we test our network architecture for multi-class classification by using the $\mathcal{D}_{TEY23}$ dataset, following the approach described in TEY23. We split the dataset into 90\% for training and 10\% for testing. As shown in Table \ref{tab:tce_catalogs}, this dataset also exhibits a strong imbalance between classes, with one predominant class, J, including astrophysical signals such as stellar variability, artifacts, and scattered light. To address this problem, we used class-weighting during training.

To perform multi-class classification, only minor changes to the architecture were required. We increased the number of output neurons from 1 to 3 and used cross-entropy as the loss function. The model was trained for 35 epochs.

The confusion matrix is given in Table \ref{tab:table_multiclass_classification}, together with the precision, recall and F1 rates for each class, and the averaged values on the entire test set. Given that our goal is to maximize the fraction of recovered planetary candidates, the class of particular interest is E, which includes planetary transits and non-contact eclipsing binaries. The fraction of correctly classified eclipsing signals is 0.67, while the rest of the samples are misclassified as J. The precision for class E is 0.587 and the recall is 0.674.

On the entire test set, the values of weighted average precision and recall are 0.776, indicating an overall good performance of the model in filtering out most non-planetary signals.

\begin{table}
\begin{tabular}{ll!{\vrule width 0pt height 10pt depth 5pt}lll|clll}
\cline{2-9}
& \textbf{Class} & \textbf{B} & \textbf{E} & \textbf{J} & \textbf{Class} & \textbf{Precision} & \textbf{Recall} & \textbf{F1-score} \\ \cline{2-9}
\multirow{4}{*}{\rotatebox{90}{\textbf{\footnotesize{Real}}}} 
& \textbf{B} & 10 (0.13) & 37 (0.50) & 26 (0.35) & \textbf{B} & 0.833 & 0.158 & 0.261 \\ 
& \textbf{E} & 0 (0.00) & 322 (0.67) & 156 (0.32) & \textbf{E} & 0.587 & 0.674 & 0.629 \\ 
& \textbf{J} & 2 (0.001) & 189 (0.15) & 1053 (0.84) & \textbf{J} & 0.860 & 0.846 & 0.853 \\ \cline{2-9}
& & \multicolumn{3}{l}{\textbf{\footnotesize{This work dispositions}}} & \textbf{Average} & 0.776 & 0.776 & 0.776 \\ \cline{6-9}
\end{tabular}
\caption{(Left block) Confusion matrix displaying the performance of our model on multi-class classification. We computed this matrix on the test set of $\mathcal{D}_{TEY23}$, corresponding to the 10\% of the entire dataset. (Right block) Precision, recall and F1-score for each class. The weighted average values computed on the entire test set are reported in the last row of the table.}
\label{tab:table_multiclass_classification}
\end{table}

\subsection{Evaluating model's performance on MES-orbital periods and MES-planet radius}
\label{sec:test_comparison_snr}
\begin{figure}
    \centering
    \includegraphics[width=\linewidth]{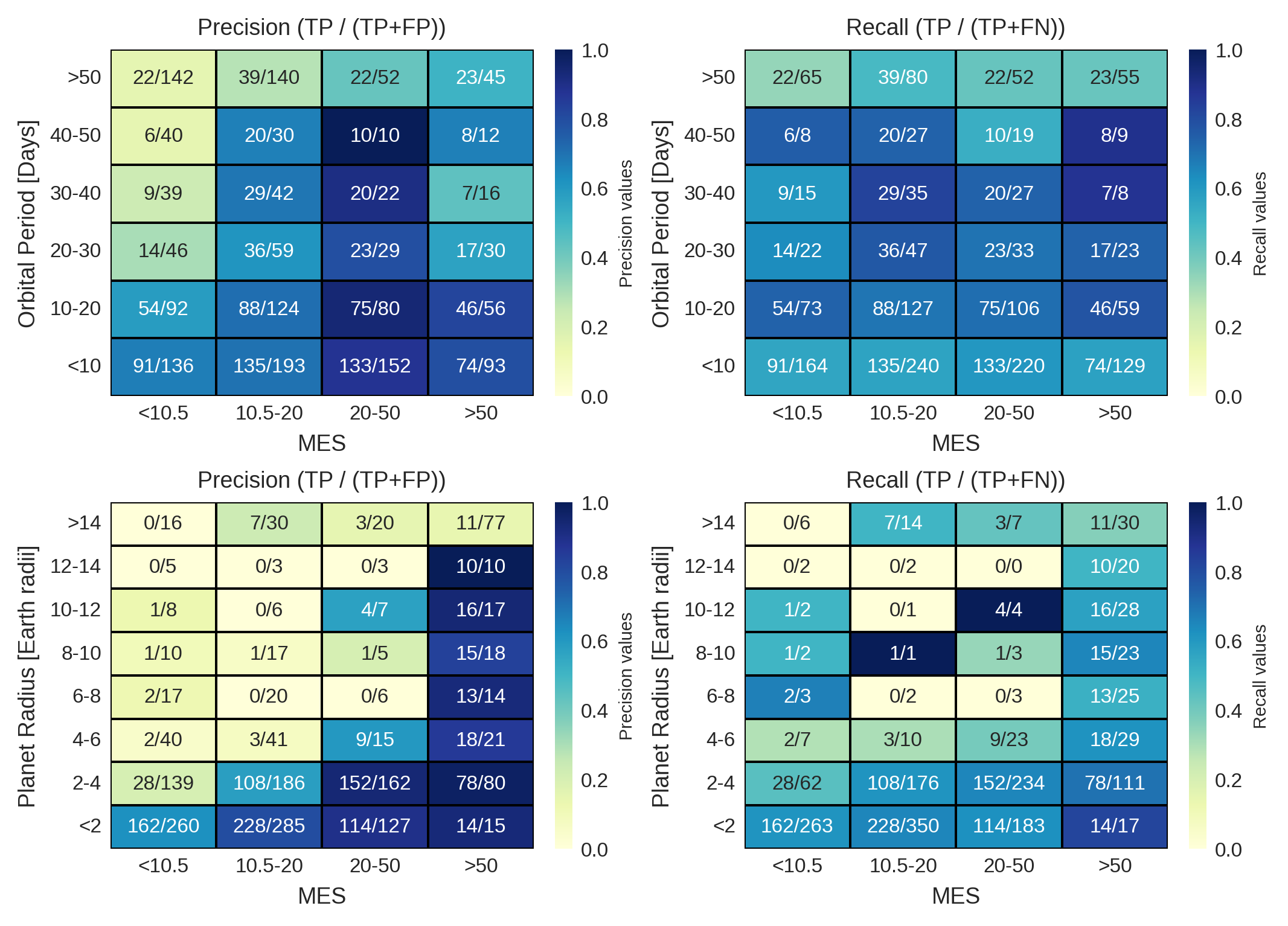}
    \caption{Performance analysis of \texttt{DART-Vetter} on the Kepler Q1-Q17 Data Release 25 dataset in the space defined by the parameters MES, Orbital Period and Planet Radius. (Top row) Fractional values of precision (left) and recall (right) as a function of MES and Orbital Period. (Bottom row) Same values obtained as a function of MES and Planet Radius.}
    \label{fig:mes_period_radius}
\end{figure}

We analyze the performance of our model as a function of the parameter pairs MES-Orbital Period and MES-Planet Radius, focusing on the dataset Kepler Q1-Q17 Data Release 25, which consists in the set of TCEs where our model struggles more in achieving robust generalization on unseen samples. The goal of this analysis is to identify those regions of the parameters space where model's performance cannot be considered reliable as precision and recall values are less than $\sim$0.60. The results discussed below are illustrated in four heatmaps in Figure \ref{fig:mes_period_radius}.

The model exhibits weak classification capabilities on TCEs with MES lower than 10.5 or with orbital period longer than 50 days. As displayed in the top left heatmap, for MES values lower than 10.5 the precision decreases from 0.66 (orbital period shorther than 10 days) to 0.15 (orbital period longer than 50 days). Conversely, for TCEs with orbital periods longer than 50 days the precision improves as the MES increases, with precision values that remain below 0.51, indicating a predominance of false positives in this region of the parameters space.

The top right heatmap shows that for TCEs with low MES ($<$10.5), our model is not able to correctly classify signals with orbital periods out of the range [10,50], where the recall assumes values lower than 0.55. These values remain very low ($<$0.48) in the region of space defined by TCEs with long orbital periods ($>$50 days), regardless an increase in MES values. In these regions, there is a higher fraction of false negatives than true positives.

Considering the bottom left heatmap, showing the performance of our model as a function of MES-Planet Radius [Earth radii, $R_{\bigoplus}$], the precision is higher than 0.85 in two particular regions: (i) for TCEs with MES larger than 50 and radius less than 14 $R_{\bigoplus}$, and (ii) for TCEs with MES values in [20,50] and radius less than 4 $R_{\bigoplus}$. Outside these regions, the model shows a marked difficulty in distinguishing planet candidates from false positives, especially for TCEs with MES lower than 10.5 or radius greater than 14 $R_{\bigoplus}$.

We observe in the bottom right heatmap a high percentage of false negatives for TCEs with MES $<$ 10.5 and radius between 2 and 6 $R_{\bigoplus}$, as well as for radius larger than 14 $R_{\bigoplus}$. While we cannot accurately assess the performance of the model on TCEs with MES $<$ 50 and radius $>$ 6 $R_{\bigoplus}$ due to a small number of samples in this region of the parameter space.

From the analysis of the four heatmaps, we find that MES is a determining factor in the correct classification of TCEs, as is orbital period. In fact, performance improves as the MES values increase, as does for orbital periods lower than 50 days. In contrast, the radius does not seem to systematically influence the performance of the model, since no trends in classification performance are identified as the radius increases. For example, in the region with MES values in the range [20-50] and radius in [10-20] the model recovers all 4  planet candidates. However, for radius up to values greater than 14 $R_{\bigoplus}$, even at high MES, the recall decreases significantly, and the model misclassifies 4 of the 7 planet candidates that fall in that parameter space.

\section{Discussion}

\label{sec:test_comparison}
Here, we compare \texttt{DART-Vetter} performance with the ones of CNNs in literature, grouped by application domain (\textit{Kepler} or TESS) and type of task (binary or multi-class). For each group of models, we pick the most effective configuration of \texttt{DART-Vetter} to perform a comparison. 

\subsection{Performance comparison with state of the art models}
\label{sec:test_comparison_comparison}
Table \ref{tab:comparison_cnn} shows for each model the scores\footnote{The scores reported in Table \ref{tab:comparison_cnn} for each model are taken from the related papers as indicated in the column Reference.} achieved on the main evaluation metrics described in Section \ref{sec:test}, and information about the related test sets (distribution of samples among the classes and source catalog). The input dimension of each model is also provided. As all the models presented in Table \ref{tab:comparison_cnn} have been trained and evaluated on slightly different distribution of samples, the comparison of their predictive performance has to be done taking this into consideration.

\begin{table}
\centering
\resizebox{0.95\columnwidth}{!}
{ \hspace{-3cm}
\begin{tabular}{lllllcll}
\hline 
\textbf{Model} & \textbf{Input size} & \textbf{Precision} & \textbf{Recall} & \textbf{F1-score} & \textbf{Test set distribution} & \textbf{Source} & \textbf{Reference} \\ \hline 
\textit{Binary classification on Kepler data} & & & & & & & \\  
Exominer & \phantom{1}1$\times$780 & \phantom{1}0.99 & \phantom{1}0.93 & \phantom{1}0.96 & \phantom{1}24 : 76 & $\mathcal{D}_{K25}$ & (1) \\ 
\textbf{Exominer++} & \phantom{1}1$\times$19,560 & \phantom{1}0.97 & \phantom{1}0.96 & \phantom{1}0.97 & \phantom{1}9 : 91 & $\mathcal{D}_{K25}$ & (2)\\ 
DART-Vetter & \phantom{1}1$\times$201 & \phantom{1}0.77 & \phantom{1}0.59 & \phantom{1}0.67 & \phantom{1}25 : 75 & $\mathcal{D}_{K25}$ & This work\\
DART-Vetter & \phantom{1}1$\times$201 & \phantom{1}0.77 & \phantom{1}0.90 & \phantom{1}0.83 & \phantom{1}37 : 63 & $\mathcal{D}_{K24}$ & This work\\ \hline
\textit{Binary classification on TESS/Kepler data} & & & & & & & \\  
Astronet-Triage & \phantom{1}1$\times$262 & \phantom{1}0.74  & \phantom{1}0.97  & \phantom{1}0.83 & \phantom{1}2 : 98 & $\mathcal{D}_{YU19}$  & (3) \\
Astronet-Vetting  & \phantom{1}1$\times$($\sim$323) & \phantom{1}0.39 & \phantom{1}0.89 & \phantom{1}0.54 & \phantom{1}2 : 98 & $\mathcal{D}_{YU19}$ & (3)\\ 
Exominer-basic & \phantom{1}1$\times$773 & \phantom{1}0.88  & \phantom{1}0.73 & \phantom{1}0.79 & \footnotesize{1167 TOIs} & $\mathcal{D}_E$ & (1) \\  
\textbf{Exominer++} & \phantom{1}1$\times$19,560 & \phantom{1}0.93 & \phantom{1}0.95 & \phantom{1}0.94 & \phantom{1}7 : 93 & TESS & (2)\\
DART-Vetter & \phantom{1}1$\times$201 & \phantom{1}0.63 & \phantom{1}0.91 & \phantom{1}0.75 & \phantom{1}30 : 70 & $\mathcal{D}_{TK}$& This work    \\ \hline
\textit{Multi-class classification on TESS data} & & & & & & & \\  
Astronet-Triage-v2 & \phantom{1}1$\times$5248 & \phantom{1}0.75 & \phantom{1}0.97 & \phantom{1}0.85 & \phantom{1}9 : 91 & $\mathcal{D}_{TEY23}$ & (4) \\
DART-Vetter & \phantom{1}1$\times$201 & \phantom{1}0.77 & \phantom{1}0.77 & \phantom{1}0.77 & 26 : 73 & $\mathcal{D}_{TEY23}$ & This work\\ \hline
\end{tabular}
}
\caption{Performance comparison of different classifiers when tested on a set of unseen Kepler/TESS TCEs. For each model we provide input size, evaluation metric scores, test set distribution, data source and corresponding references. Input sizes were calculated by flattening all input vectors and supplementary features processed by the models. \textbf{References:} (1) \citet{2022ApJ...926..120V}; (2) \citet{valizadegan2025exominer}; (3) \citet{2019AJ....158...25Y}; (4) \citet{2023AJ....165...95T}.}
\label{tab:comparison_cnn}
\end{table}

In the binary classification of \textit{Kepler} TCEs, \texttt{DART-Vetter} shows discrete but inferior performance than \texttt{Exominer}, with an F1-score 13\% lower than the CNN developed by \citet{2022ApJ...926..120V}. Yet, with a recall of 0.90 with respect to 0.93 of \texttt{Exominer}, \texttt{DART-Vetter} shows comparable triage capabilities. 
But the input dimensionality\footnote{We computed all the input sizes described here and reported in Table \ref{tab:comparison_cnn} by flattening the entire set of input vectors and supplementary features processed by the models.} of our model is significantly reduced (1$\times$201 vs 1$\times$780), indicating a better computational efficiency and generalization on cross-mission data.

For binary classification on TESS TCEs, the best-performing CNN currently available to our knowledge is still \texttt{Astronet-Triage} by \citet{2019AJ....158...25Y}. 
By working solely on the global and local views, this model achieves appreciable performance on the test set, with a precision of 74\% and a recall of 97\%, when including eclipsing binaries in the true positive class. \texttt{Astronet-Vetting} is a slightly more complex model with respect to \texttt{Astronet-Triage} as it processes a larger number of input features, but achieves lower performance, supporting the finding by  \citet{2022A&C....4100654V} and subsequently by this study: an increase in input dimensionality does not always yield benefits in terms of model's performance. Both versions of Astronet have been trained and tested on TESS TCEs only. On the other hand, \texttt{Exominer-basic} is trained on TCEs from the Kepler Q1-Q17 Data Release 25 catalog \citep{2016AJ....152..158T} and tested on 1,167 TESS TCEs for which \citet{2021ApJS..254...39G} provided dispositions. This test set is composed of TCEs labelled as KP, CP, false positives and false alarms. The authors deliberately decided to not include the TCEs dispositioned as PC as they state the following: "We did not include the TOIs with PC disposition because this disposition is not conclusive". As the number of test samples for each class is not detailed in \citet{2022ApJ...926..120V}, we only report the total test set size in Table \ref{tab:comparison_cnn}.
\texttt{Exominer-basic} reaches noteworthy performance by working on time-series (including the global view) and stellar parameters thus showing that such CNNs are able to perform well on cross-mission TCEs \citep[see also][]{fiscale_wirn}. In this context, \texttt{DART-Vetter} stands out as the model that needs the lower input dimensionality. Its triage capabilities are comparable with those of \texttt{Astronet-Triage} but evaluated on a more carefully balanced dataset.

We include also the comparison with the improvement of the model provided by Valizadegan et al (2022), Exominer++ \citep{valizadegan2025exominer}. This work, available on arXiv, has been developed to improve performance on TESS data. The authors extend the architecture of this CNN by introducing five additional input information. The model processes as input photometric data for a total dimension of 19,522, and 38 scalar features. In the binary classification on \textit{Kepler} data, \texttt{Exominer++} improves its previous F1-score by 1\%, establishing itself as the best model currently presented in this context. Predictive performance improves significantly on  TESS data. As we found with \texttt{DART-Vetter}, \texttt{Exominer++} also improves its generalization capabilities on TESS when trained on a dataset combining \textit{Kepler} and TESS data. With a recall of 95\%, the model is inferior only to \texttt{Astronet-Triage} in the correct classification of planets, while it proves to be the best in terms of generalization with an F1-score of 94\%. We calculated the distributions of the test sets on which \texttt{Exominer++} is evaluated based on the number of TCEs given in Tables 2 and 4 of their paper, and considering that the model is trained and tested with 10-fold cross-validation.

In multi-class classification, the F1-score of \texttt{DART-Vetter} is 8\% lower than that of the optimal model \texttt{Astronet-Triage-v2}. Even in this case the model stands out for its extremely compact input, using only 1$\times$201 features compared with 1$\times$5,248 of Astronet-Triage-v2.

Thus, the minimized complexity of our model ensures it can be adapted on binary or multi-class classification of signals of any domain, with triage capabilities comparable to those of the currently available models specific to a given transit survey. However, it is important to consider that such a reduction in complexity, due in part to the use of input photometric data alone, limits the potential of our model, which shows degrading performances in classifying \textit{Kepler} TCEs with MES $<$ 20 or with orbital periods $>$ 50 days.

On the other hand, the weakness of state-of-the-art models lies in the high computational complexity required by the processes of (i) data collection, (ii) pre-processing, and (iii) network architecture implementation. With \texttt{DART-Vetter} we strive to achieve a good trade-off between predictive performance and ease of implementing. 

\subsection{Rationale behind choosing a less complex model}
\label{sec:test_comparison_rationale}

   Minimizing model complexity is a fundamental aspect of the \texttt{DART-Vetter} design, as it enhances training efficiency, improves generalization to unseen data, and increases interpretability. A cornerstone in this area is the seminal work by \citet{guyon2003introduction}. Their work addresses the importance of variable and feature selection, which allows for processing only linearly independent features, thereby eliminating redundant information in the data that adds unnecessary complexity to the problem. This is a very relevant topic in Deep Learning and recent studies are continuing to focus on it \citep{im2022computational,pudjihartono2022review}.
    
    As extensively described in \citet{hu2021model}, in the context of Deep Learning, model complexity refers to different factors (e.g. expressive capacity, effective model complexity). For \texttt{DART-Vetter}, we consider the following set of factors: the number of weights to be optimized during the training process; the number of hyperparameters to be set during the fine-tuning process; the computational complexity required by the process of preparing the input dataset; and the flexibility of the model in terms of application on data from different surveys. Models with a large number of weights need to be trained on very large datasets to avoid overfitting and underfitting. In an overfitting scenario, the model stores such specific details about the training set that it cannot generalize to unseen data. In contrast, in an underfitting scenario, which generally occurs for very complex models trained for a few epochs on a relatively small dataset, the model is unable to learn from the training data the features necessary to generalize over the test set. Occam's razor principle suggests that, among models that achieve similar performance, the simplest model (with the least number of weights) should be preferred, since this choice minimizes the risk of overfitting and underfitting.

As described in Section \ref{sec:model_architecture}, the use of a single classification layer dramatically reduces the number of network parameters (or weights), leading to a crucial advantage: keep model complexity low while retaining high predictive performance.
    
    With 527,329 parameters, \texttt{DART-Vetter} is significantly smaller than more complex models such as \texttt{Astronet-Triage-v2}, which has over 100 million parameters. Despite this reduction in complexity, \texttt{DART-Vetter} achieves competitive F1-scores on different domains, demonstrating that a less complex network can achieve performance comparable to that of larger models, as previously shown by \citet{2022JAI....1150011V}. This architectural simplification reduces the risks of overfitting and underfitting by improving the ability of the model to generalize on unseen data. It also limits the need for extensive hyperparameters fine-tuning or significant architectural modifications to deal with datasets from different missions, such as \textit{Kepler}, TESS, or PLATO. Indeed, although models such as \texttt{Exominer} and \texttt{Astronet-Triage-v2} are highly effective on TCEs belonging to the same domain as their respective training sets, their application on data from different missions would require substantial and computationally expensive architectural changes. An example of this is \texttt{Exominer-basic}, the version of \texttt{Exominer} adapted for TESS data. This adaptation from \textit{Kepler} to TESS results in a decay of model performance by about 17\% in terms of F1-score. These changes not only require an almost complete redesign of the network architecture, but also significantly compromise model performance. 

    With \texttt{DART-Vetter}, the reduction in the number of trainable weights results in greater computational efficiency, both during training and in the preparation of inputs. In particular, high-dimensional inputs result in a larger number of parameters to be optimized and a high cost for dataset generation, as data from different sources need to be consistently combined. 

    Thus, our approach not only reduces computational costs but also makes the model inherently flexible in terms of adaptability to data from different missions, ensuring effective, efficient, and portable triage.

\section{Current limitation and future perspectives}
\label{sec:limitations}

Here we address the current factors that constrain the predictive accuracy of our model with the aim of improving its performance and interpretability in future applications \citep{Maratea2021}.

\subsection{Representing the transit shape} 
The global views we generated for each TCE consist in light curves that are binned and phase-folded over the TCEs periods. In this work, we found the same problem discussed in \citet{2023AJ....165...95T}. More precisely, binning and phase-folding operations involve aggregating flux measurements into discrete bins, leading to a reduction in the overall information content. Although these processes aid in the generation of a more compact input representation, they introduce uncertainties since they degrade the precision of the representation of the transit shape. In this study, we did not investigate about the quantification of the impact resulting from the loss of information during phase folding and binning. This issue remains a matter of interest and a particular focus will be given in future works.

\subsection{The bias of TCEs dispositions} 
\label{sec:limitations_dispositions}
Our CNN for exoplanet detection is inevitably affected by biases introduced by human-generated labels, and its predictions lack explainability. During the vetting process, human vetters could inadvertently introduce their own biases or subjective judgments. This can results in a biased training dataset and subsequently affects the performance of our model trained on such data. 
To address this issue, it is necessary to employ Deep Learning models specifically designed to learn from both labels and intrinsic patterns in the data. Among these models, we plan to use Graph Neural Networks \citep[GNNs;][]{scarselli2008graph, wu2020comprehensive}. Transforming an $m$-dimensional input vector (i.e. light curve) into a graph, with the aim of processing it by means of a GNN, could allow us to detect more complex non-linear relationships and patterns within the input\citep{prummel2023inductive}. 
Consequently, the use of GNNs might enhances the model's interpretability and provide insights into patterns contributing to the discrimination of exoplanets from false positives.

\section{Conclusions}
\label{sec:conclusion}

The CNNs for exoplanets detection have demonstrated remarkable predictive performance, as shown by the works of \citet{2022ApJ...926..120V}, \citet{2023AJ....165...95T} and \citet{valizadegan2025exominer}, while also maintaining simplicity in implementation, as seen in the study conducted by \citet{2022A&C....4100654V}. Building upon these achievements, we developed an effective and easily replicable CNN to automatically distinguish planetary transits from false positives. 

Starting from the promising outcomes obtained in our previous work \citep{fiscale_wirn}, we deployed \texttt{DART-Vetter}, a triage model capable of performing both binary and multi-class classification on TCEs from different surveys. Our results emphasize the ability of the model in solving these tasks when assessed on single- (\textit{Kepler} and TESS) and cross-domain (TESS + Kepler DR24) datasets, achieving recall values on test sets greater than 92\%. 
On the other side, our model does not achieve good performance on Kepler DR25 TCEs, with a high fraction of false negatives of 40\%. As shown in Section \ref{sec:test_comparison_snr}, \texttt{DART-Vetter} struggles with the classification of \textit{Kepler} TCEs orbiting their stars in more than 50 days, or with MES lower than 20.

By comparing \texttt{DART-Vetter} with other CNNs for exoplanets detection we found that the performance are lower in terms of absolute accuracy with respect to state-of-the-art models such as \texttt{Exominer}, \texttt{Exominer++} and \texttt{Astronet-Triage-v2}. But the strength of our model lies in its straightforward adaptability across binary/multi-class classification on several datasets. This is accomplished thanks to the reduced model complexity, which we obtained mainly by minimizing the input dimensionality and the number of classification layers, together with the relative number of neurons. 
Unlike top performing models, which require high computational complexity for data collection, pre-processing and network implementation, \texttt{DART-Vetter} offers an optimal balance between predictive performance and implementation simplicity, minimizing the false negatives fraction for planets with short orbital periods (except for the TCEs of $\mathcal{D}_{K25}$ where more in-depth evaluations have been presented in Section \ref{sec:test_comparison_snr}) by needing only period-folded and binned light curves. \\

This paper includes data collected by the Kepler mission and obtained from the MAST data archive at the Space Telescope Science Institute (STScI). Funding for the Kepler mission is provided by the NASA Science Mission Directorate. STScI is operated by the Association of Universities for Research in Astronomy, Inc., under NASA contract NAS 5–26555.

This paper includes data collected with the TESS mission, obtained from the MAST data archive at the Space Telescope Science Institute (STScI). Funding for the TESS mission is provided by the NASA Explorer Program. STScI is operated by the Association of Universities for Research in Astronomy, Inc., under NASA contract NAS 5–26555.

This research has made use of the Exoplanet Follow-up Observation Program (ExoFOP; DOI: 10.26134/ExoFOP5) website, which is operated by the California Institute of Technology, under contract with the National Aeronautics and Space Administration under the Exoplanet Exploration Program.

This research has made use of the NASA Exoplanet Archive, which is operated by the California Institute of Technology, under contract with the National Aeronautics and Space Administration under the Exoplanet Exploration Program.

\appendix
Appendix \ref{sec:appendix_a} provides the details related to the model selection phase, while in Appendix \ref{sec:appendix_b} we show how model performance vary when training and testing on different dataset splits.

\section{Appendix A}
\label{sec:appendix_a}
The model selection phase is aimed at determining the best performing network on the test set and it involves training and evaluating multiple models with varying architectures and hyperparameters (e.g., learning rate, batch size). We conducted this phase on the Kepler DR24 dataset as the one characterized by a more balanced number of positive and negative TCEs, with a discrete S/N when compared to other sources of signals. The hyperparameters and the relative grids of values are listed in Table \ref{tab:model_selection_hyperparameters}, where the values in bold represent the sub-optimal configuration returned by the grid search. In Table \ref{tab:model_selection_hyperparameters}, the training epochs increase by 10 in the interval [10,100], and its optimal value depends on the input dataset. The upper limit of fully-connected neurons was set at 768 as it corresponds to the output size of the feature extraction branch. The dropout rate is evaluated on a discrete grid in [0.2, 0.5] as this is the interval suggested in \citet{srivastava}. The underlined configuration is defined as sub-optimal as the fixed grids assume values in finite and discrete intervals.

We began with the base network architecture designed by \citet{fiscale_wirn}. This architecture processes two views of the same TCE: the global view and local view, that is a zoomed-in representation of the first. To create a simplified architecture, denoted as $\mathcal{M}$, we initially (i) removed the branch processing the local view and (ii) reduced the size of the convolutional filters from $5\times1$ to $3\times1$. We removed the local view because it is obtained from the global view using a different binning size. Thus, the local view is linearly dependent with respect to the global view. In \citet{guyon2003introduction} it was shown that processing linearly dependent input data adds further complexity to the problem but no additional information is gained by considering them. We adopted the use of smaller convolutional filters after \citet{ciresan2011flexible} and \citet{2014arXiv1409.1556S} proved their effectiveness in capturing more complex relationships in the input data than larger filters. The use of smaller convolutional window exactly suits our needs since the most relevant information in our input samples (the transit points) might also be represented by a single point in the global view (see Figure 3 of \citet{2018AJ....155...94S}).

The model selection process was then conducted in two steps:
\begin{enumerate}
    \item [] \textbf{STEP 1}. Based on $\mathcal{M}$, we trained a set of models by varying their depth (from five to three convolutional blocks) and evaluating the use of batch normalization layers within the convolutional and classification blocks;
    \item [] \textbf{STEP 2}. We modified each convolutional block by removing one of the two triplets (1D convolution layer - Rectified Linear Unit - Dropout) and repeated the \textit{STEP 1} for one iteration.
\end{enumerate}
This methodology led to the architecture of the model presented in this study, \texttt{DART-Vetter}, characterized by $\sim$1,400,000 parameters less to be optimized during the training process. The differences in network architecture, between the base model proposed in \citet{fiscale_wirn} and \texttt{DART-Vetter}, obtained through model selection are summarized in Table \ref{tab:model_selection_before_after}.

\begin{table}[]
    \centering
\begin{tabular}{ll} \hline
        \textbf{Hyperparameter} & \textbf{Values} \\ \hline 
        Activation function & \textbf{ReLU}, Leaky ReLU\\
        Batch size & 64, \textbf{128}, 256, 512, 768, 1024\\
        Dropout rate & \textbf{0.2}, 0.3, 0.4, 0.5\\
        Training epochs & 10:10:100, 150, 200, 300, 400, 500\\
        Fully-connected neurons & 256, \textbf{512}, 768\\
        Layer weights initializer & \begin{tabular}[c]{@{}l@{}}{}\hspace{-1cm}Uniform, \textbf{Lecun Uniform},  Random normal,\\ \hspace{-1cm}Zero, Glorot Normal, Glorot Uniform,\\ \hspace{-1cm}He Normal, He Uniform{}\end{tabular} \\
        Learning rate & 1e-2, \textbf{1e-3}, 1e-4, 1e-5\\
        Optimizer & \textbf{Adam}\\ \hline 
    \end{tabular}
\caption{List of hyperparameters and their respective values on which we based the grid search process. For each hyperparameter, a set of discrete values is defined. The grid search process trained the model based on all possible combinations among the distinct values that each hyperparameter can assume. We highlight in bold the values we used to train the best model.}
    \label{tab:model_selection_hyperparameters}
\end{table}

\begin{table}[]
    \centering
    \begin{tabular}{lll}
    \hline 
         & Fiscale et al. (2023) & This work\\ \hline
        \textit{Input} & & \\
        \# Views & 2 & 1\\ \hline 
        \textit{Feature Extraction branch} &  & \\
        \# Convolutional blocks & 5+2 & 5\\
         Convolutional filters & 5$\times$1 & 3$\times$1\\ \hline 
         \textit{Classification branch}&  & \\
        \# Dense blocks & 1 & 1\\
        \# Neurons & 768 & 512\\
         Batch normalization & yes & no\\ \hline
        \textit{Number of parameters} &  & \\
        Feature Extraction branch & $\sim$670,000 & $\sim$127,000\\
        Classification branch & $\sim$1,260,000 & $\sim$400,000\\ 
        Total & $\sim$1,927,000 & $\sim$527,000\\ \hline 
    \end{tabular}
    \caption{Summary of the characteristics of the model before \citep{fiscale_wirn} and after (This work) the model selection phase. The element denoted with the '+' symbol corresponds to the values for the two branches processing the global and local views.}
    \label{tab:model_selection_before_after}
\end{table}

\section{Appendix B}
\label{sec:appendix_b}

\begin{figure}[H]
    \centering
    \includegraphics[width=\linewidth]{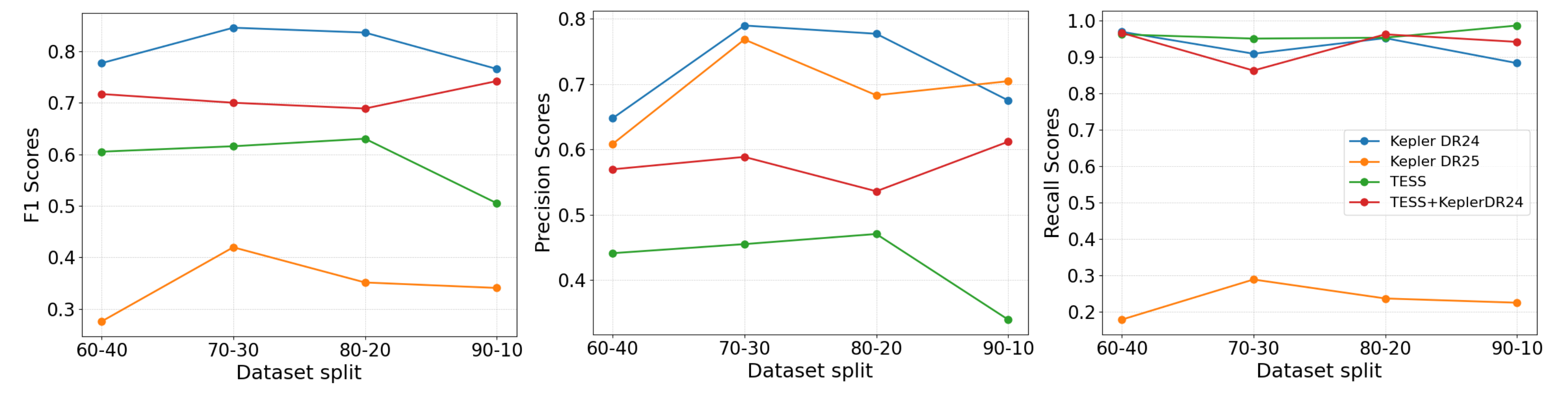}
    \caption{Variation of the main evaluation metrics as the training and test datasets are split according to the different following ratios: 60-40, 70-30, 80-20 and 90-10. Each line shows the scores of F1 (left panel), precision (middle panel) and recall (right panel) for a given dataset.} 
    \label{fig:performance_over_datasplit}
\end{figure}

As depicted in Figure \ref{fig:performance_over_datasplit}, the values of precision, recall, and F1-score computed on each test set do not exhibit significant changes when varying the data splits used for training and testing the model. This outcome highlights the ability of \texttt{DART-Vetter} in guaranteeing optimal performance on the dataset under analysis, even when the training set is relatively small. Such a stability in performance across different training and test set fractions can be mainly attributed to the reduced number of trainable parameters of the model.
Overall, optimal performance in terms of F1-score are achieved when training-testing the model on the data split 70-30.

\bibliography{new.ms.bib}{}

\begin{thebibliography}{}
\expandafter\ifx\csname natexlab\endcsname\relax\def\natexlab#1{#1}\fi
\providecommand{\url}[1]{\href{#1}{#1}}
\providecommand{\dodoi}[1]{doi:~\href{http://doi.org/#1}{\nolinkurl{#1}}}
\providecommand{\doeprint}[1]{\href{http://ascl.net/#1}{\nolinkurl{http://ascl.net/#1}}}
\providecommand{\doarXiv}[1]{\href{https://arxiv.org/abs/#1}{\nolinkurl{https://arxiv.org/abs/#1}}}

\bibitem[{{Akeson} \& {Christiansen}(2019)}]{2019AAS...23314009A}
{Akeson}, R., \& {Christiansen}, J. 2019, in American Astronomical Society
  Meeting Abstracts, Vol. 233, American Astronomical Society Meeting Abstracts
  \#233, 140.09

\bibitem[{{Ansdell} {et~al.}(2018){Ansdell}, {Ioannou}, {Osborn}, {Sasdelli},
  {2018 NASA Frontier Development Lab Exoplanet Team}, {Smith}, {Caldwell},
  {Jenkins}, {R{\"a}issi}, {Angerhausen}, \& {NASA Frontier Development Lab
  Exoplanet Mentors}}]{2018ApJ...869L...7A}
{Ansdell}, M., {Ioannou}, Y., {Osborn}, H.~P., {et~al.} 2018, \apjl, 869, L7,
  \dodoi{10.3847/2041-8213/aaf23b}

\bibitem[{{Armstrong} {et~al.}(2018){Armstrong}, {G{\"u}nther}, {McCormac},
  {Smith}, {Bayliss}, {Bouchy}, {Burleigh}, {Casewell}, {Eigm{\"u}ller},
  {Gillen}, {Goad}, {Hodgkin}, {Jenkins}, {Louden}, {Metrailler}, {Pollacco},
  {Poppenhaeger}, {Queloz}, {Raynard}, {Rauer}, {Udry}, {Walker}, {Watson},
  {West}, \& {Wheatley}}]{2018MNRAS.478.4225A}
{Armstrong}, D.~J., {G{\"u}nther}, M.~N., {McCormac}, J., {et~al.} 2018,
  \mnras, 478, 4225, \dodoi{10.1093/mnras/sty1313}

\bibitem[{{Bryson} {et~al.}(2010){Bryson}, {Tenenbaum}, {Jenkins},
  {Chandrasekaran}, {Klaus}, {Caldwell}, {Gilliland}, {Haas}, {Dotson}, {Koch},
  \& {Borucki}}]{2010ApJ...713L..97B}
{Bryson}, S.~T., {Tenenbaum}, P., {Jenkins}, J.~M., {et~al.} 2010, \apjl, 713,
  L97, \dodoi{10.1088/2041-8205/713/2/L97}

\bibitem[{Bryson {et~al.}(2013)Bryson, Jenkins, Gilliland, Twicken, Clarke,
  Rowe, Caldwell, Batalha, Mullally, Haas, {et~al.}}]{bryson2013identification}
Bryson, S.~T., Jenkins, J.~M., Gilliland, R.~L., {et~al.} 2013, Publications of
  the Astronomical Society of the Pacific, 125, 889

\bibitem[{{Cacciapuoti} {et~al.}(2022){Cacciapuoti}, {Kostov}, {Kuchner},
  {Quintana}, {Col{\'o}n}, {Brande}, {Mullally}, {Chance}, {Christiansen},
  {Ahlers}, {Di Fraia}, {Durantini Luca}, {Ienco}, {Gallo}, {de Lima}, {Hyogo},
  {Andr{\'e}s-Carcasona}, {Fornear}, {de Lambilly}, {Salik}, {Yablonsky},
  {Wallace}, \& {Acharya}}]{2022MNRAS.513..102C}
{Cacciapuoti}, L., {Kostov}, V.~B., {Kuchner}, M., {et~al.} 2022, \mnras, 513,
  102, \dodoi{10.1093/mnras/stac652}

\bibitem[{Catanzarite(2015)}]{catanzarite2015autovetter}
Catanzarite, J.~H. 2015, Astronomy \& Astrophysics

\bibitem[{{Chaushev} {et~al.}(2019){Chaushev}, {Raynard}, {Goad},
  {Eigm{\"u}ller}, {Armstrong}, {Briegal}, {Burleigh}, {Casewell}, {Gill},
  {Jenkins}, {Nielsen}, {Watson}, {West}, {Wheatley}, {Udry}, \&
  {Vines}}]{2019MNRAS.488.5232C}
{Chaushev}, A., {Raynard}, L., {Goad}, M.~R., {et~al.} 2019, \mnras, 488, 5232,
  \dodoi{10.1093/mnras/stz2058}

\bibitem[{Christiansen {et~al.}(2012)Christiansen, Jenkins, Caldwell, Barclay,
  Bryson, Burke, Clarke, Coughlin, Girouard, Haas, Hall, Ibrahim, Klaus,
  Kolodziejczak, Li, McCauliff, Morris, Mullally, Quintana, Rowe, Sabale,
  Sanderfer, Seader, Smith, Still, Tenenbaum, Thompson, Twicken, \&
  Uddin}]{christiansenkepler}
Christiansen, J.~L., Jenkins, J.~M., Caldwell, D.~A., {et~al.} 2012

\bibitem[{Ciresan {et~al.}(2011)Ciresan, Meier, Masci, Gambardella, \&
  Schmidhuber}]{ciresan2011flexible}
Ciresan, D.~C., Meier, U., Masci, J., Gambardella, L.~M., \& Schmidhuber, J.
  2011, in Twenty-second international joint conference on artificial
  intelligence, Citeseer

\bibitem[{Claeskens {et~al.}(2008)Claeskens, Hjort,
  {et~al.}}]{claeskens2008model}
Claeskens, G., Hjort, N.~L., {et~al.} 2008, Model selection and model
  averaging, Vol. 330 (Cambridge University Press Cambridge)

\bibitem[{Coughlin {et~al.}(2016)Coughlin, Mullally, Thompson, Rowe, Burke,
  Latham, Batalha, Ofir, Quarles, Henze, Wolfgang, Caldwell, Bryson, Shporer,
  Catanzarite, Akeson, Barclay, Borucki, Boyajian, Campbell, Christiansen,
  Girouard, Haas, Howell, Huber, Jenkins, Li, Patil-Sabale, Quintana, Ramirez,
  Seader, Smith, Tenenbaum, Twicken, \& Zamudio}]{Coughlin_2016}
Coughlin, J.~L., Mullally, F., Thompson, S.~E., {et~al.} 2016, The
  Astrophysical Journal Supplement Series, 224, 12,
  \dodoi{10.3847/0067-0049/224/1/12}

\bibitem[{Cybenko(1989)}]{cybenko1989approximation}
Cybenko, G. 1989, Mathematics of control, signals and systems, 2, 303

\bibitem[{{Dattilo} {et~al.}(2019){Dattilo}, {Vanderburg}, {Shallue}, {Mayo},
  {Berlind}, {Bieryla}, {Calkins}, {Esquerdo}, {Everett}, {Howell}, {Latham},
  {Scott}, \& {Yu}}]{2019AJ....157..169D}
{Dattilo}, A., {Vanderburg}, A., {Shallue}, C.~J., {et~al.} 2019, \aj, 157,
  169, \dodoi{10.3847/1538-3881/ab0e12}

\bibitem[{De~Luca {et~al.}(2019)De~Luca, Fiscale, Landolfi, \&
  Di~Mauro}]{de2019distributed}
De~Luca, P., Fiscale, S., Landolfi, L., \& Di~Mauro, A. 2019, in Internet and
  Distributed Computing Systems: 12th International Conference, IDCS 2019,
  Naples, Italy, October 10--12, 2019, Proceedings 12, Springer, 369--378

\bibitem[{De~Luca {et~al.}(2022)De~Luca, Galletti, \& Marcellino}]{10090063}
De~Luca, P., Galletti, A., \& Marcellino, L. 2022, in 2022 16th International
  Conference on Signal-Image Technology \& Internet-Based Systems (SITIS),
  530--534, \dodoi{10.1109/SITIS57111.2022.00085}

\bibitem[{Durairaj \& Mohan(2022)}]{durairaj2022convolutional}
Durairaj, D.~M., \& Mohan, B.~K. 2022, Neural Computing and Applications, 34,
  13319

\bibitem[{Ferone \& Petrosino(2017)}]{Ferone2017116}
Ferone, A., \& Petrosino, A. 2017, Lecture Notes in Computer Science (including
  subseries Lecture Notes in Artificial Intelligence and Lecture Notes in
  Bioinformatics), 10147 LNAI, 116 – 125,
  \dodoi{10.1007/978-3-319-52962-2_10}

\bibitem[{Fiscale {et~al.}(2025)Fiscale, Ferone, Ciaramella, Inno,
  Giordano~Orsini, Covone, \& Rotundi}]{fiscale2025detection}
Fiscale, S., Ferone, A., Ciaramella, A., {et~al.} 2025, Electronics, 14, 1738

\bibitem[{Fiscale {et~al.}(2021)Fiscale, Luca, Inno, Marcellino, Galletti,
  Rotundi, Ciaramella, Covone, \& Quintana}]{fiscale2021gpu}
Fiscale, S., Luca, P.~D., Inno, L., {et~al.} 2021, in International Conference
  on Computational Science, Springer, 420--432

\bibitem[{Fiscale {et~al.}(2023)Fiscale, Inno, Ciaramella, Ferone, Rotundi,
  De~Luca, Galletti, Marcellino, \& Covone}]{fiscale_wirn}
Fiscale, S., Inno, L., Ciaramella, A., {et~al.} 2023, in Applications of
  Artificial Intelligence and Neural Systems to Data Science (Springer),
  127--135

\bibitem[{{Gaia Collaboration} {et~al.}(2018){Gaia Collaboration}, {Babusiaux},
  {van Leeuwen}, {Barstow}, {Jordi}, {Vallenari}, {Bossini}, {Bressan},
  {Cantat-Gaudin}, {van Leeuwen}, {Brown}, {Prusti}, {de Bruijne},
  {Bailer-Jones}, {Biermann}, {Evans}, {Eyer}, {Jansen}, {Klioner}, {Lammers},
  {Lindegren}, {Luri}, {Mignard}, {Panem}, {Pourbaix}, {Randich}, {Sartoretti},
  {Siddiqui}, {Soubiran}, {Walton}, {Arenou}, {Bastian}, {Cropper}, {Drimmel},
  {Katz}, {Lattanzi}, {Bakker}, {Cacciari}, {Casta{\~n}eda}, {Chaoul}, {Cheek},
  {De Angeli}, {Fabricius}, {Guerra}, {Holl}, {Masana}, {Messineo}, {Mowlavi},
  {Nienartowicz}, {Panuzzo}, {Portell}, {Riello}, {Seabroke}, {Tanga},
  {Th{\'e}venin}, {Gracia-Abril}, {Comoretto}, {Garcia-Reinaldos}, {Teyssier},
  {Altmann}, {Andrae}, {Audard}, {Bellas-Velidis}, {Benson}, {Berthier},
  {Blomme}, {Burgess}, {Busso}, {Carry}, {Cellino}, {Clementini}, {Clotet},
  {Creevey}, {Davidson}, {De Ridder}, {Delchambre}, {Dell'Oro}, {Ducourant},
  {Fern{\'a}ndez-Hern{\'a}ndez}, {Fouesneau}, {Fr{\'e}mat}, {Galluccio},
  {Garc{\'\i}a-Torres}, {Gonz{\'a}lez-N{\'u}{\~n}ez}, {Gonz{\'a}lez-Vidal},
  {Gosset}, {Guy}, {Halbwachs}, {Hambly}, {Harrison}, {Hern{\'a}ndez},
  {Hestroffer}, {Hodgkin}, {Hutton}, {Jasniewicz}, {Jean-Antoine-Piccolo},
  {Jordan}, {Korn}, {Krone-Martins}, {Lanzafame}, {Lebzelter}, {L{\"o}ffler},
  {Manteiga}, {Marrese}, {Mart{\'\i}n-Fleitas}, {Moitinho}, {Mora}, {Muinonen},
  {Osinde}, {Pancino}, {Pauwels}, {Petit}, {Recio-Blanco}, {Richards},
  {Rimoldini}, {Robin}, {Sarro}, {Siopis}, {Smith}, {Sozzetti}, {S{\"u}veges},
  {Torra}, {van Reeven}, {Abbas}, {Abreu Aramburu}, {Accart}, {Aerts},
  {Altavilla}, {{\'A}lvarez}, {Alvarez}, {Alves}, {Anderson}, {Andrei},
  {Anglada Varela}, {Antiche}, {Antoja}, {Arcay}, {Astraatmadja}, {Bach},
  {Baker}, {Balaguer-N{\'u}{\~n}ez}, {Balm}, {Barache}, {Barata}, {Barbato},
  {Barblan}, {Barklem}, {Barrado}, {Barros}, {Bartholom{\'e} Mu{\~n}oz},
  {Bassilana}, {Becciani}, {Bellazzini}, {Berihuete}, {Bertone}, {Bianchi},
  {Bienaym{\'e}}, {Blanco-Cuaresma}, {Boch}, {Boeche}, {Bombrun}, {Borrachero},
  {Bouquillon}, {Bourda}, {Bragaglia}, {Bramante}, {Breddels}, {Brouillet},
  {Br{\"u}semeister}, {Brugaletta}, {Bucciarelli}, {Burlacu}, {Busonero},
  {Butkevich}, {Buzzi}, {Caffau}, {Cancelliere}, {Cannizzaro}, {Carballo},
  {Carlucci}, {Carrasco}, {Casamiquela}, {Castellani}, {Castro-Ginard},
  {Charlot}, {Chemin}, {Chiavassa}, {Cocozza}, {Costigan}, {Cowell}, {Crifo},
  {Crosta}, {Crowley}, {Cuypers}, {Dafonte}, {Damerdji}, {Dapergolas}, {David},
  {David}, {de Laverny}, {De Luise}, {De March}, {de Martino}, {de Souza}, {de
  Torres}, {Debosscher}, {del Pozo}, {Delbo}, {Delgado}, {Delgado}, {Diakite},
  {Diener}, {Distefano}, {Dolding}, {Drazinos}, {Dur{\'a}n}, {Edvardsson},
  {Enke}, {Eriksson}, {Esquej}, {Eynard Bontemps}, {Fabre}, {Fabrizio},
  {Faigler}, {Falc{\~a}o}, {Farr{\`a}s Casas}, {Federici}, {Fedorets},
  {Fernique}, {Figueras}, {Filippi}, {Findeisen}, {Fonti}, {Fraile}, {Fraser},
  {Fr{\'e}zouls}, {Gai}, {Galleti}, {Garabato}, {Garc{\'\i}a-Sedano},
  {Garofalo}, {Garralda}, {Gavel}, {Gavras}, {Gerssen}, {Geyer}, {Giacobbe},
  {Gilmore}, {Girona}, {Giuffrida}, {Glass}, {Gomes}, {Granvik}, {Gueguen},
  {Guerrier}, {Guiraud}, {Guti{\'e}}, {Haigron}, {Hatzidimitriou}, {Hauser},
  {Haywood}, {Heiter}, {Helmi}, {Heu}, {Hilger}, {Hobbs}, {Hofmann}, {Holland},
  {Huckle}, {Hypki}, {Icardi}, {Jan{\ss}en}, {Jevardat de Fombelle}, {Jonker},
  {Juh{\'a}sz}, {Julbe}, {Karampelas}, {Kewley}, {Klar}, {Kochoska}, {Kohley},
  {Kolenberg}, {Kontizas}, {Kontizas}, {Koposov}, {Kordopatis},
  {Kostrzewa-Rutkowska}, {Koubsky}, {Lambert}, {Lanza}, {Lasne}, {Lavigne}, {Le
  Fustec}, {Le Poncin-Lafitte}, {Lebreton}, {Leccia}, {Leclerc},
  {Lecoeur-Taibi}, {Lenhardt}, {Leroux}, {Liao}, {Licata}, {Lindstr{\o}m},
  {Lister}, {Livanou}, {Lobel}, {L{\'o}pez}, {Managau}, {Mann}, {Mantelet},
  {Marchal}, {Marchant}, {Marconi}, {Marinoni}, {Marschalk{\'o}}, {Marshall},
  {Martino}, {Marton}, {Mary}, {Massari}, {Matijevi{\v{c}}}, {Mazeh},
  {McMillan}, {Messina}, {Michalik}, {Millar}, {Molina}, {Molinaro},
  {Moln{\'a}r}, {Montegriffo}, {Mor}, {Morbidelli}, {Morel}, {Morris},
  {Mulone}, {Muraveva}, {Musella}, {Nelemans}, {Nicastro}, {Noval},
  {O'Mullane}, {Ord{\'e}novic}, {Ord{\'o}{\~n}ez-Blanco}, {Osborne}, {Pagani},
  {Pagano}, {Pailler}, {Palacin}, {Palaversa}, {Panahi}, {Pawlak},
  {Piersimoni}, {Pineau}, {Plachy}, {Plum}, {Poggio}, {Poujoulet},
  {Pr{\v{s}}a}, {Pulone}, {Racero}, {Ragaini}, {Rambaux}, {Ramos-Lerate},
  {Regibo}, {Reyl{\'e}}, {Riclet}, {Ripepi}, {Riva}, {Rivard}, {Rixon},
  {Roegiers}, {Roelens}, {Romero-G{\'o}mez}, {Rowell}, {Royer}, {Ruiz-Dern},
  {Sadowski}, {Sagrist{\`a} Sell{\'e}s}, {Sahlmann}, {Salgado}, {Salguero},
  {Sanna}, {Santana-Ros}, {Sarasso}, {Savietto}, {Schultheis}, {Sciacca},
  {Segol}, {Segovia}, {S{\'e}gransan}, {Shih}, {Siltala}, {Silva}, {Smart},
  {Smith}, {Solano}, {Solitro}, {Sordo}, {Soria Nieto}, {Souchay}, {Spagna},
  {Spoto}, {Stampa}, {Steele}, {Steidelm{\"u}ller}, {Stephenson}, {Stoev},
  {Suess}, {Surdej}, {Szabados}, {Szegedi-Elek}, {Tapiador}, {Taris}, {Tauran},
  {Taylor}, {Teixeira}, {Terrett}, {Teyssandier}, {Thuillot}, {Titarenko},
  {Torra Clotet}, {Turon}, {Ulla}, {Utrilla}, {Uzzi}, {Vaillant}, {Valentini},
  {Valette}, {van Elteren}, {Van Hemelryck}, {Vaschetto}, {Vecchiato},
  {Veljanoski}, {Viala}, {Vicente}, {Vogt}, {von Essen}, {Voss}, {Votruba},
  {Voutsinas}, {Walmsley}, {Weiler}, {Wertz}, {Wevers}, {Wyrzykowski},
  {Yoldas}, {{\v{Z}}erjal}, {Ziaeepour}, {Zorec}, {Zschocke}, {Zucker},
  {Zurbach}, \& {Zwitter}}]{2018A&A...616A..10G}
{Gaia Collaboration}, {Babusiaux}, C., {van Leeuwen}, F., {et~al.} 2018, \aap,
  616, A10, \dodoi{10.1051/0004-6361/201832843}

\bibitem[{{Giacalone} {et~al.}(2021){Giacalone}, {Dressing}, {Jensen},
  {Collins}, {Ricker}, {Vanderspek}, {Seager}, {Winn}, {Jenkins}, {Barclay},
  {Barkaoui}, {Cadieux}, {Charbonneau}, {Collins}, {Conti}, {Doyon}, {Evans},
  {Ghachoui}, {Gillon}, {Guerrero}, {Hart}, {Jehin}, {Kielkopf}, {McLean},
  {Murgas}, {Palle}, {Parviainen}, {Pozuelos}, {Relles}, {Shporer}, {Socia},
  {Stockdale}, {Tan}, {Torres}, {Twicken}, {Waalkes}, \&
  {Waite}}]{2021AJ....161...24G}
{Giacalone}, S., {Dressing}, C.~D., {Jensen}, E. L.~N., {et~al.} 2021, \aj,
  161, 24, \dodoi{10.3847/1538-3881/abc6af}

\bibitem[{{Guerrero} {et~al.}(2021){Guerrero}, {Seager}, {Huang}, {Vanderburg},
  {Garcia Soto}, {Mireles}, {Hesse}, {Fong}, {Glidden}, {Shporer}, {Latham},
  {Collins}, {Quinn}, {Burt}, {Dragomir}, {Crossfield}, {Vanderspek},
  {Fausnaugh}, {Burke}, {Ricker}, {Daylan}, {Essack}, {G{\"u}nther}, {Osborn},
  {Pepper}, {Rowden}, {Sha}, {Villanueva}, {Yahalomi}, {Yu}, {Ballard},
  {Batalha}, {Berardo}, {Chontos}, {Dittmann}, {Esquerdo}, {Mikal-Evans},
  {Jayaraman}, {Krishnamurthy}, {Louie}, {Mehrle}, {Niraula}, {Rackham},
  {Rodriguez}, {Rowden}, {Sousa-Silva}, {Watanabe}, {Wong}, {Zhan},
  {Zivanovic}, {Christiansen}, {Ciardi}, {Swain}, {Lund}, {Mullally},
  {Fleming}, {Rodriguez}, {Boyd}, {Quintana}, {Barclay}, {Col{\'o}n},
  {Rinehart}, {Schlieder}, {Clampin}, {Jenkins}, {Twicken}, {Caldwell},
  {Coughlin}, {Henze}, {Lissauer}, {Morris}, {Rose}, {Smith}, {Tenenbaum},
  {Ting}, {Wohler}, {Bakos}, {Bean}, {Berta-Thompson}, {Bieryla}, {Bouma},
  {Buchhave}, {Butler}, {Charbonneau}, {Doty}, {Ge}, {Holman}, {Howard},
  {Kaltenegger}, {Kane}, {Kjeldsen}, {Kreidberg}, {Lin}, {Minsky}, {Narita},
  {Paegert}, {P{\'a}l}, {Palle}, {Sasselov}, {Spencer}, {Sozzetti}, {Stassun},
  {Torres}, {Udry}, \& {Winn}}]{2021ApJS..254...39G}
{Guerrero}, N.~M., {Seager}, S., {Huang}, C.~X., {et~al.} 2021, \apjs, 254, 39,
  \dodoi{10.3847/1538-4365/abefe1}

\bibitem[{Guyon \& Elisseeff(2003)}]{guyon2003introduction}
Guyon, I., \& Elisseeff, A. 2003, Journal of machine learning research, 3, 1157

\bibitem[{Hadjigeorghiou \& Armstrong(2024)}]{hadjigeorghiou2024positional}
Hadjigeorghiou, A., \& Armstrong, D.~J. 2024, Monthly Notices of the Royal
  Astronomical Society, 527, 4018

\bibitem[{Hornik {et~al.}(1989)Hornik, Stinchcombe, \&
  White}]{hornik1989multilayer}
Hornik, K., Stinchcombe, M., \& White, H. 1989, Neural networks, 2, 359

\bibitem[{{Howell} {et~al.}(2014){Howell}, {Sobeck}, {Haas}, {Still},
  {Barclay}, {Mullally}, {Troeltzsch}, {Aigrain}, {Bryson}, {Caldwell},
  {Chaplin}, {Cochran}, {Huber}, {Marcy}, {Miglio}, {Najita}, {Smith},
  {Twicken}, \& {Fortney}}]{2014PASP..126..398H}
{Howell}, S.~B., {Sobeck}, C., {Haas}, M., {et~al.} 2014, \pasp, 126, 398,
  \dodoi{10.1086/676406}

\bibitem[{Hu {et~al.}(2021)Hu, Chu, Pei, Liu, \& Bian}]{hu2021model}
Hu, X., Chu, L., Pei, J., Liu, W., \& Bian, J. 2021, Knowledge and Information
  Systems, 63, 2585

\bibitem[{{Huang}(2020)}]{10.17909/t9-r086-e880}
{Huang}, C.~X. 2020, TESS Lightcurves From The MIT Quick-Look Pipeline ("QLP"),
   STScI/MAST, \dodoi{10.17909/t9-r086-e880}

\bibitem[{{Huang} {et~al.}(2020{\natexlab{a}}){Huang}, {Vanderburg}, {P{\'a}l},
  {Sha}, {Yu}, {Fong}, {Fausnaugh}, {Shporer}, {Guerrero}, {Vanderspek}, \&
  {Ricker}}]{2020RNAAS...4..204H}
{Huang}, C.~X., {Vanderburg}, A., {P{\'a}l}, A., {et~al.} 2020{\natexlab{a}},
  Research Notes of the American Astronomical Society, 4, 204,
  \dodoi{10.3847/2515-5172/abca2e}

\bibitem[{{Huang} {et~al.}(2020{\natexlab{b}}){Huang}, {Vanderburg}, {P{\'a}l},
  {Sha}, {Yu}, {Fong}, {Fausnaugh}, {Shporer}, {Guerrero}, {Vanderspek}, \&
  {Ricker}}]{2020RNAAS...4..206H}
---. 2020{\natexlab{b}}, Research Notes of the American Astronomical Society,
  4, 206, \dodoi{10.3847/2515-5172/abca2d}

\bibitem[{Im \& Dasari(2022)}]{im2022computational}
Im, M.~S., \& Dasari, V.~R. 2022, arXiv preprint arXiv:2207.14620

\bibitem[{Ioffe \& Szegedy(2015)}]{ioffe2015batch}
Ioffe, S., \& Szegedy, C. 2015, in International conference on machine
  learning, PMLR, 448--456

\bibitem[{Jara-Maldonado {et~al.}(2020)Jara-Maldonado, Alarcon-Aquino, \&
  Rosas-Romero}]{jara2020multiresolution}
Jara-Maldonado, M., Alarcon-Aquino, V., \& Rosas-Romero, R. 2020, in Mexican
  International Conference on Artificial Intelligence, Springer, 50--64

\bibitem[{{Jenkins} {et~al.}(2002){Jenkins}, {Caldwell}, \&
  {Borucki}}]{2002ApJ...564..495J}
{Jenkins}, J.~M., {Caldwell}, D.~A., \& {Borucki}, W.~J. 2002, \apj, 564, 495,
  \dodoi{10.1086/324143}

\bibitem[{Jenkins {et~al.}(2010)Jenkins, Caldwell, Chandrasekaran, Twicken,
  Bryson, Quintana, Clarke, Li, Allen, Tenenbaum,
  {et~al.}}]{jenkins2010overview}
Jenkins, J.~M., Caldwell, D.~A., Chandrasekaran, H., {et~al.} 2010, The
  Astrophysical Journal Letters, 713, L87

\bibitem[{{Jenkins} {et~al.}(2010){Jenkins}, {Chandrasekaran}, {McCauliff},
  {Caldwell}, {Tenenbaum}, {Li}, {Klaus}, {Cote}, \&
  {Middour}}]{2010SPIE.7740E..0DJ}
{Jenkins}, J.~M., {Chandrasekaran}, H., {McCauliff}, S.~D., {et~al.} 2010, in
  Society of Photo-Optical Instrumentation Engineers (SPIE) Conference Series,
  Vol. 7740, Software and Cyberinfrastructure for Astronomy, ed. N.~M.
  {Radziwill} \& A.~{Bridger}, 77400D, \dodoi{10.1117/12.856764}

\bibitem[{Jenkins {et~al.}(2016)Jenkins, Twicken, McCauliff, Campbell,
  Sanderfer, Lung, Mansouri-Samani, Girouard, Tenenbaum, Klaus,
  {et~al.}}]{jenkins2016tess}
Jenkins, J.~M., Twicken, J.~D., McCauliff, S., {et~al.} 2016, in Software and
  Cyberinfrastructure for Astronomy IV, Vol. 9913, SPIE, 1232--1251

\bibitem[{Kingma \& Ba(2015)}]{kingma2014adam}
Kingma, D.~P., \& Ba, J. 2015, in 3rd International Conference on Learning
  Representations, ICLR 2015, San Diego, CA, USA, May 7-9, 2015.

\bibitem[{Klambauer {et~al.}(2017)Klambauer, Unterthiner, Mayr, \&
  Hochreiter}]{klambauer2017self}
Klambauer, G., Unterthiner, T., Mayr, A., \& Hochreiter, S. 2017, Advances in
  neural information processing systems, 30

\bibitem[{{Koch} {et~al.}(2010){Koch}, {Borucki}, {Basri}, {Batalha}, {Brown},
  {Caldwell}, {Christensen-Dalsgaard}, {Cochran}, {DeVore}, {Dunham},
  {Gautier}, {Geary}, {Gilliland}, {Gould}, {Jenkins}, {Kondo}, {Latham},
  {Lissauer}, {Marcy}, {Monet}, {Sasselov}, {Boss}, {Brownlee}, {Caldwell},
  {Dupree}, {Howell}, {Kjeldsen}, {Meibom}, {Morrison}, {Owen}, {Reitsema},
  {Tarter}, {Bryson}, {Dotson}, {Gazis}, {Haas}, {Kolodziejczak}, {Rowe}, {Van
  Cleve}, {Allen}, {Chandrasekaran}, {Clarke}, {Li}, {Quintana}, {Tenenbaum},
  {Twicken}, \& {Wu}}]{2010ApJ...713L..79K}
{Koch}, D.~G., {Borucki}, W.~J., {Basri}, G., {et~al.} 2010, \apjl, 713, L79,
  \dodoi{10.1088/2041-8205/713/2/L79}

\bibitem[{{Kostov} {et~al.}(2019){Kostov}, {Mullally}, {Quintana}, {Coughlin},
  {Mullally}, {Barclay}, {Col{\'o}n}, {Schlieder}, {Barentsen}, \&
  {Burke}}]{2019AJ....157..124K}
{Kostov}, V.~B., {Mullally}, S.~E., {Quintana}, E.~V., {et~al.} 2019, \aj, 157,
  124, \dodoi{10.3847/1538-3881/ab0110}

\bibitem[{{Kov{\'a}cs} {et~al.}(2002){Kov{\'a}cs}, {Zucker}, \&
  {Mazeh}}]{2002A&A...391..369K}
{Kov{\'a}cs}, G., {Zucker}, S., \& {Mazeh}, T. 2002, \aap, 391, 369,
  \dodoi{10.1051/0004-6361:20020802}

\bibitem[{{Kunimoto} {et~al.}(2022){Kunimoto}, {Tey}, {Fong}, {Hesse},
  {Shporer}, {Fausnaugh}, {Vanderspek}, \& {Ricker}}]{2022RNAAS...6..236K}
{Kunimoto}, M., {Tey}, E., {Fong}, W., {et~al.} 2022, Research Notes of the
  American Astronomical Society, 6, 236, \dodoi{10.3847/2515-5172/aca158}

\bibitem[{{Kunimoto} {et~al.}(2021){Kunimoto}, {Huang}, {Tey}, {Fong}, {Hesse},
  {Shporer}, {Guerrero}, {Fausnaugh}, {Vanderspek}, \&
  {Ricker}}]{2021RNAAS...5..234K}
{Kunimoto}, M., {Huang}, C., {Tey}, E., {et~al.} 2021, Research Notes of the
  American Astronomical Society, 5, 234, \dodoi{10.3847/2515-5172/ac2ef0}

\bibitem[{LeCun {et~al.}(2015)LeCun, Bengio, \& Hinton}]{lecun2015deep}
LeCun, Y., Bengio, Y., \& Hinton, G. 2015, nature, 521, 436

\bibitem[{Lertnattee \& Theeramunkong(2004)}]{lertnattee2004analysis}
Lertnattee, V., \& Theeramunkong, T. 2004, in IEEE International Symposium on
  Communications and Information Technology, 2004. ISCIT 2004., Vol.~2, IEEE,
  1171--1176

\bibitem[{{Liao} {et~al.}(2024){Liao}, {Ren}, {Chen}, {Li}, \&
  {Li}}]{2024AJ....167..180L}
{Liao}, H., {Ren}, G., {Chen}, X., {Li}, Y., \& {Li}, G. 2024, \aj, 167, 180,
  \dodoi{10.3847/1538-3881/ad298f}

\bibitem[{Magliano {et~al.}(2022)Magliano, Covone, Dobal, Cacciapuoti,
  Tonietti, Giacalone, Vines, Inno, Jenkins, Lissauer, Bieryla, Oliva, Pagano,
  Kostov, Ziegler, Ciardi, Gonzales, Dressing, Buchhave, Howell, Matson,
  Matthews, Rotundi, Alves, Fiscale, Ienco, Peña, Gallo, \&
  Muscari~Tomajoli}]{10.1093/mnras/stac3404}
Magliano, C., Covone, G., Dobal, R., {et~al.} 2022, Monthly Notices of the
  Royal Astronomical Society, 519, 1562, \dodoi{10.1093/mnras/stac3404}

\bibitem[{{Magliano} {et~al.}(2023){Magliano}, {Kostov}, {Cacciapuoti},
  {Covone}, {Inno}, {Fiscale}, {Kuchner}, {Quintana}, {Salik}, {Saggese},
  {Yablonsky}, {Fornear}, {Hyogo}, {Di Fraia}, {Luca}, {de Lambilly}, {Oliva},
  {Pagano}, {Ienco}, {de Lima}, {Andr{\'e}s-Carcasona}, {Gallo}, \&
  {Acharya}}]{2023MNRAS.521.3749M}
{Magliano}, C., {Kostov}, V., {Cacciapuoti}, L., {et~al.} 2023, \mnras, 521,
  3749, \dodoi{10.1093/mnras/stad683}

\bibitem[{Maratea \& Ferone(2021)}]{Maratea2021}
Maratea, A., \& Ferone, A. 2021, CEUR Workshop Proceedings, 3074

\bibitem[{{MAST Team}(2021)}]{10.17909/t9-nmc8-f686}
{MAST Team}. 2021, TESS Light Curves - All Sectors,  STScI/MAST,
  \dodoi{10.17909/t9-nmc8-f686}

\bibitem[{{McCauliff} {et~al.}(2015){McCauliff}, {Jenkins}, {Catanzarite},
  {Burke}, {Coughlin}, {Twicken}, {Tenenbaum}, {Seader}, {Li}, \&
  {Cote}}]{2015ApJ...806....6M}
{McCauliff}, S.~D., {Jenkins}, J.~M., {Catanzarite}, J., {et~al.} 2015, \apj,
  806, 6, \dodoi{10.1088/0004-637X/806/1/6}

\bibitem[{{Mislis} {et~al.}(2016){Mislis}, {Bachelet}, {Alsubai}, {Bramich}, \&
  {Parley}}]{2016MNRAS.455..626M}
{Mislis}, D., {Bachelet}, E., {Alsubai}, K.~A., {Bramich}, D.~M., \& {Parley},
  N. 2016, \mnras, 455, 626, \dodoi{10.1093/mnras/stv2333}

\bibitem[{{Morris} {et~al.}(2020){Morris}, {Twicken}, {Smith}, {Clarke},
  {Jenkins}, {Bryson}, {Girouard}, \& {Klaus}}]{2020ksci.rept....6M}
{Morris}, R.~L., {Twicken}, J.~D., {Smith}, J.~C., {et~al.} 2020, {Kepler Data
  Processing Handbook: Photometric Analysis}, Kepler Science Document
  KSCI-19081-003, id. 6. Edited by Jon M. Jenkins.

\bibitem[{{Morton}(2015)}]{2015ascl.soft03011M}
{Morton}, T.~D. 2015, {VESPA: False positive probabilities calculator},
  Astrophysics Source Code Library, record ascl:1503.011.
\newblock \doeprint{1503.011}

\bibitem[{{Morton} {et~al.}(2016){Morton}, {Bryson}, {Coughlin}, {Rowe},
  {Ravichandran}, {Petigura}, {Haas}, \& {Batalha}}]{2016ApJ...822...86M}
{Morton}, T.~D., {Bryson}, S.~T., {Coughlin}, J.~L., {et~al.} 2016, \apj, 822,
  86, \dodoi{10.3847/0004-637X/822/2/86}

\bibitem[{{Mullally} {et~al.}(2016){Mullally}, {Coughlin}, {Thompson},
  {Christiansen}, {Burke}, {Clarke}, \& {Haas}}]{2016PASP..128g4502M}
{Mullally}, F., {Coughlin}, J.~L., {Thompson}, S.~E., {et~al.} 2016, \pasp,
  128, 074502, \dodoi{10.1088/1538-3873/128/965/074502}

\bibitem[{Orsini {et~al.}(2025)Orsini, Ferone, Inno, Giacobbe, Maratea,
  Ciaramella, Bonomo, \& Rotundi}]{orsini2025data}
Orsini, M.~G., Ferone, A., Inno, L., {et~al.} 2025, Astronomy and Computing,
  100964

\bibitem[{{Prummel} {et~al.}(2023){Prummel}, {Giraldo}, {Zakharova}, \&
  {Bouwmans}}]{prummel2023inductive}
{Prummel}, W., {Giraldo}, J.~H., {Zakharova}, A., \& {Bouwmans}, T. 2023, arXiv
  e-prints, arXiv:2305.09585, \dodoi{10.48550/arXiv.2305.09585}

\bibitem[{Pudjihartono {et~al.}(2022)Pudjihartono, Fadason, Kempa-Liehr, \&
  O'Sullivan}]{pudjihartono2022review}
Pudjihartono, N., Fadason, T., Kempa-Liehr, A.~W., \& O'Sullivan, J.~M. 2022,
  Frontiers in Bioinformatics, 2, 927312

\bibitem[{Ribani \& Marengoni(2019)}]{ribani2019survey}
Ribani, R., \& Marengoni, M. 2019, in 2019 32nd SIBGRAPI conference on
  graphics, patterns and images tutorials (SIBGRAPI-T), IEEE, 47--57

\bibitem[{{Ricker} {et~al.}(2014){Ricker}, {Winn}, {Vanderspek}, {Latham},
  {Bakos}, {Bean}, {Berta-Thompson}, {Brown}, {Buchhave}, {Butler}, {Butler},
  {Chaplin}, {Charbonneau}, {Christensen-Dalsgaard}, {Clampin}, {Deming},
  {Doty}, {De Lee}, {Dressing}, {Dunham}, {Endl}, {Fressin}, {Ge}, {Henning},
  {Holman}, {Howard}, {Ida}, {Jenkins}, {Jernigan}, {Johnson}, {Kaltenegger},
  {Kawai}, {Kjeldsen}, {Laughlin}, {Levine}, {Lin}, {Lissauer}, {MacQueen},
  {Marcy}, {McCullough}, {Morton}, {Narita}, {Paegert}, {Palle}, {Pepe},
  {Pepper}, {Quirrenbach}, {Rinehart}, {Sasselov}, {Sato}, {Seager},
  {Sozzetti}, {Stassun}, {Sullivan}, {Szentgyorgyi}, {Torres}, {Udry}, \&
  {Villasenor}}]{2014SPIE.9143E..20R}
{Ricker}, G.~R., {Winn}, J.~N., {Vanderspek}, R., {et~al.} 2014, in Society of
  Photo-Optical Instrumentation Engineers (SPIE) Conference Series, Vol. 9143,
  Space Telescopes and Instrumentation 2014: Optical, Infrared, and Millimeter
  Wave, ed. J.~{Oschmann}, Jacobus~M., M.~{Clampin}, G.~G. {Fazio}, \& H.~A.
  {MacEwen}, 914320, \dodoi{10.1117/12.2063489}

\bibitem[{Scarselli {et~al.}(2008)Scarselli, Gori, Tsoi, Hagenbuchner, \&
  Monfardini}]{scarselli2008graph}
Scarselli, F., Gori, M., Tsoi, A.~C., Hagenbuchner, M., \& Monfardini, G. 2008,
  IEEE transactions on neural networks, 20, 61

\bibitem[{{Schanche} {et~al.}(2019){Schanche}, {Collier Cameron},
  {H{\'e}brard}, {Nielsen}, {Triaud}, {Almenara}, {Alsubai}, {Anderson},
  {Armstrong}, {Barros}, {Bouchy}, {Boumis}, {Brown}, {Faedi}, {Hay}, {Hebb},
  {Kiefer}, {Mancini}, {Maxted}, {Palle}, {Pollacco}, {Queloz}, {Smalley},
  {Udry}, {West}, \& {Wheatley}}]{2019MNRAS.483.5534S}
{Schanche}, N., {Collier Cameron}, A., {H{\'e}brard}, G., {et~al.} 2019,
  \mnras, 483, 5534, \dodoi{10.1093/mnras/sty3146}

\bibitem[{{Seader} {et~al.}(2015){Seader}, {Jenkins}, {Tenenbaum}, {Twicken},
  {Smith}, {Morris}, {Catanzarite}, {Clarke}, {Li}, {Cote}, {Burke},
  {McCauliff}, {Girouard}, {Campbell}, {Kamal Uddin}, {Zamudio}, {Sabale},
  {Henze}, {Thompson}, \& {Klaus}}]{2015ApJS..217...18S}
{Seader}, S., {Jenkins}, J.~M., {Tenenbaum}, P., {et~al.} 2015, \apjs, 217, 18,
  \dodoi{10.1088/0067-0049/217/1/18}

\bibitem[{{Shallue} \& {Vanderburg}(2018)}]{2018AJ....155...94S}
{Shallue}, C.~J., \& {Vanderburg}, A. 2018, \aj, 155, 94,
  \dodoi{10.3847/1538-3881/aa9e09}

\bibitem[{{Simonyan} \& {Zisserman}(2014)}]{2014arXiv1409.1556S}
{Simonyan}, K., \& {Zisserman}, A. 2014, arXiv e-prints, arXiv:1409.1556,
  \dodoi{10.48550/arXiv.1409.1556}

\bibitem[{{Smith} {et~al.}(2012){Smith}, {Stumpe}, {Van Cleve}, {Jenkins},
  {Barclay}, {Fanelli}, {Girouard}, {Kolodziejczak}, {McCauliff}, {Morris}, \&
  {Twicken}}]{2012PASP..124.1000S}
{Smith}, J.~C., {Stumpe}, M.~C., {Van Cleve}, J.~E., {et~al.} 2012, \pasp, 124,
  1000, \dodoi{10.1086/667697}

\bibitem[{Srivastava {et~al.}(2014)Srivastava, Hinton, Krizhevsky, Sutskever,
  \& Salakhutdinov}]{srivastava}
Srivastava, N., Hinton, G., Krizhevsky, A., Sutskever, I., \& Salakhutdinov, R.
  2014, The journal of machine learning research, 15, 1929

\bibitem[{{Stassun} {et~al.}(2019){Stassun}, {Oelkers}, {Paegert}, {Torres},
  {Pepper}, {De Lee}, {Collins}, {Latham}, {Muirhead}, {Chittidi},
  {Rojas-Ayala}, {Fleming}, {Rose}, {Tenenbaum}, {Ting}, {Kane}, {Barclay},
  {Bean}, {Brassuer}, {Charbonneau}, {Ge}, {Lissauer}, {Mann}, {McLean},
  {Mullally}, {Narita}, {Plavchan}, {Ricker}, {Sasselov}, {Seager}, {Sharma},
  {Shiao}, {Sozzetti}, {Stello}, {Vanderspek}, {Wallace}, \&
  {Winn}}]{2019AJ....158..138S}
{Stassun}, K.~G., {Oelkers}, R.~J., {Paegert}, M., {et~al.} 2019, \aj, 158,
  138, \dodoi{10.3847/1538-3881/ab3467}

\bibitem[{{STScI}(2016)}]{10.17909/T98304}
{STScI}. 2016, Kepler LC+SC, Q0-Q17,  STScI/MAST, \dodoi{10.17909/T98304}

\bibitem[{Szegedy {et~al.}(2015)Szegedy, Liu, Jia, Sermanet, Reed, Anguelov,
  Erhan, Vanhoucke, \& Rabinovich}]{szegedy2015going}
Szegedy, C., Liu, W., Jia, Y., {et~al.} 2015, in Proceedings of the IEEE
  conference on computer vision and pattern recognition, 1--9

\bibitem[{{Tey} {et~al.}(2023){Tey}, {Moldovan}, {Kunimoto}, {Huang},
  {Shporer}, {Daylan}, {Muthukrishna}, {Vanderburg}, {Dattilo}, {Ricker}, \&
  {Seager}}]{2023AJ....165...95T}
{Tey}, E., {Moldovan}, D., {Kunimoto}, M., {et~al.} 2023, \aj, 165, 95,
  \dodoi{10.3847/1538-3881/acad85}

\bibitem[{{Thompson} {et~al.}(2018){Thompson}, {Coughlin}, {Hoffman},
  {Mullally}, {Christiansen}, {Burke}, {Bryson}, {Batalha}, {Haas},
  {Catanzarite}, {Rowe}, {Barentsen}, {Caldwell}, {Clarke}, {Jenkins}, {Li},
  {Latham}, {Lissauer}, {Mathur}, {Morris}, {Seader}, {Smith}, {Klaus},
  {Twicken}, {Van Cleve}, {Wohler}, {Akeson}, {Ciardi}, {Cochran}, {Henze},
  {Howell}, {Huber}, {Pr{\v{s}}a}, {Ram{\'\i}rez}, {Morton}, {Barclay},
  {Campbell}, {Chaplin}, {Charbonneau}, {Christensen-Dalsgaard}, {Dotson},
  {Doyle}, {Dunham}, {Dupree}, {Ford}, {Geary}, {Girouard}, {Isaacson},
  {Kjeldsen}, {Quintana}, {Ragozzine}, {Shabram}, {Shporer}, {Silva Aguirre},
  {Steffen}, {Still}, {Tenenbaum}, {Welsh}, {Wolfgang}, {Zamudio}, {Koch}, \&
  {Borucki}}]{2018ApJS..235...38T}
{Thompson}, S.~E., {Coughlin}, J.~L., {Hoffman}, K., {et~al.} 2018, \apjs, 235,
  38, \dodoi{10.3847/1538-4365/aab4f9}

\bibitem[{{Tonry} {et~al.}(2018){Tonry}, {Denneau}, {Flewelling}, {Heinze},
  {Onken}, {Smartt}, {Stalder}, {Weiland}, \& {Wolf}}]{2018ApJ...867..105T}
{Tonry}, J.~L., {Denneau}, L., {Flewelling}, H., {et~al.} 2018, \apj, 867, 105,
  \dodoi{10.3847/1538-4357/aae386}

\bibitem[{{Twicken} {et~al.}(2010){Twicken}, {Chandrasekaran}, {Jenkins},
  {Gunter}, {Girouard}, \& {Klaus}}]{2010SPIE.7740E..1UT}
{Twicken}, J.~D., {Chandrasekaran}, H., {Jenkins}, J.~M., {et~al.} 2010, in
  Society of Photo-Optical Instrumentation Engineers (SPIE) Conference Series,
  Vol. 7740, Software and Cyberinfrastructure for Astronomy, ed. N.~M.
  {Radziwill} \& A.~{Bridger}, 77401U, \dodoi{10.1117/12.856798}

\bibitem[{{Twicken} {et~al.}(2016){Twicken}, {Jenkins}, {Seader}, {Tenenbaum},
  {Smith}, {Brownston}, {Burke}, {Catanzarite}, {Clarke}, {Cote}, {Girouard},
  {Klaus}, {Li}, {McCauliff}, {Morris}, {Wohler}, {Campbell}, {Kamal Uddin},
  {Zamudio}, {Sabale}, {Bryson}, {Caldwell}, {Christiansen}, {Coughlin},
  {Haas}, {Henze}, {Sanderfer}, \& {Thompson}}]{2016AJ....152..158T}
{Twicken}, J.~D., {Jenkins}, J.~M., {Seader}, S.~E., {et~al.} 2016, \aj, 152,
  158, \dodoi{10.3847/0004-6256/152/6/158}

\bibitem[{{Valizadegan} {et~al.}(2023){Valizadegan}, {Martinho}, {Jenkins},
  {Caldwell}, {Twicken}, \& {Bryson}}]{2023AJ....166...28V}
{Valizadegan}, H., {Martinho}, M. J.~S., {Jenkins}, J.~M., {et~al.} 2023, \aj,
  166, 28, \dodoi{10.3847/1538-3881/acd344}

\bibitem[{{Valizadegan} {et~al.}(2022){Valizadegan}, {Martinho}, {Wilkens},
  {Jenkins}, {Smith}, {Caldwell}, {Twicken}, {Gerum}, {Walia}, {Hausknecht},
  {Lubin}, {Bryson}, \& {Oza}}]{2022ApJ...926..120V}
{Valizadegan}, H., {Martinho}, M. J.~S., {Wilkens}, L.~S., {et~al.} 2022, \apj,
  926, 120, \dodoi{10.3847/1538-4357/ac4399}

\bibitem[{Valizadegan {et~al.}(2025)Valizadegan, Martinho, Jenkins, Twicken,
  Caldwell, Maynard, Wei, Zhong, Yates, Donald,
  {et~al.}}]{valizadegan2025exominer}
Valizadegan, H., Martinho, M.~J., Jenkins, J.~M., {et~al.} 2025, arXiv preprint
  arXiv:2502.09790

\bibitem[{{Vanderburg} \& {Johnson}(2014)}]{2014PASP..126..948V}
{Vanderburg}, A., \& {Johnson}, J.~A. 2014, \pasp, 126, 948,
  \dodoi{10.1086/678764}

\bibitem[{{Visser} {et~al.}(2022{\natexlab{a}}){Visser}, {Bosma}, \&
  {Postma}}]{2022JAI....1150011V}
{Visser}, K., {Bosma}, B., \& {Postma}, E. 2022{\natexlab{a}}, Journal of
  Astronomical Instrumentation, 11, 2250011, \dodoi{10.1142/S2251171722500118}

\bibitem[{{Visser} {et~al.}(2022{\natexlab{b}}){Visser}, {Bosma}, \&
  {Postma}}]{2022A&C....4100654V}
---. 2022{\natexlab{b}}, Astronomy and Computing, 41, 100654,
  \dodoi{10.1016/j.ascom.2022.100654}

\bibitem[{Wu {et~al.}(2020)Wu, Pan, Chen, Long, Zhang, \&
  Philip}]{wu2020comprehensive}
Wu, Z., Pan, S., Chen, F., {et~al.} 2020, IEEE transactions on neural networks
  and learning systems, 32, 4

\bibitem[{{Yu} {et~al.}(2019){Yu}, {Vanderburg}, {Huang}, {Shallue},
  {Crossfield}, {Gaudi}, {Daylan}, {Dattilo}, {Armstrong}, {Ricker},
  {Vanderspek}, {Latham}, {Seager}, {Dittmann}, {Doty}, {Glidden}, \&
  {Quinn}}]{2019AJ....158...25Y}
{Yu}, L., {Vanderburg}, A., {Huang}, C., {et~al.} 2019, \aj, 158, 25,
  \dodoi{10.3847/1538-3881/ab21d6}

\bibitem[{{Zacharias} {et~al.}(2017){Zacharias}, {Finch}, \&
  {Frouard}}]{2017AJ....153..166Z}
{Zacharias}, N., {Finch}, C., \& {Frouard}, J. 2017, \aj, 153, 166,
  \dodoi{10.3847/1538-3881/aa6196}

\bibitem[{{Zacharias} {et~al.}(2013){Zacharias}, {Finch}, {Girard}, {Henden},
  {Bartlett}, {Monet}, \& {Zacharias}}]{2013AJ....145...44Z}
{Zacharias}, N., {Finch}, C.~T., {Girard}, T.~M., {et~al.} 2013, \aj, 145, 44,
  \dodoi{10.1088/0004-6256/145/2/44}

\bibitem[{Zhao {et~al.}(2017)Zhao, Lu, Chen, Liu, \&
  Wu}]{zhao2017convolutional}
Zhao, B., Lu, H., Chen, S., Liu, J., \& Wu, D. 2017, Journal of Systems
  Engineering and Electronics, 28, 162

\end{thebibliography}
\bibliographystyle{aasjournal}

\end{document}